\documentclass[12pt]{article}

\usepackage[utf8]{inputenc}
\usepackage[T1]{fontenc}
\usepackage{amsmath,amssymb}

\usepackage{newtxtext,newtxmath}
\usepackage[margin=1in]{geometry}
\usepackage{ragged2e}
\usepackage{graphicx}
\usepackage{booktabs}
\usepackage{hyperref}
\hypersetup{hidelinks}
\usepackage{xcolor}
\usepackage{algorithm}
\usepackage{algpseudocode}
\usepackage{natbib}
\usepackage{longtable}
\usepackage{array}
\usepackage{caption}
\usepackage{etoolbox}
\usepackage[section]{placeins}
\usepackage{float}
\usepackage{needspace}
\title{Measuring Self-Rating Bias in LLM-Generated Survey Data: A Semantic Similarity Framework for Independent Scale Mapping}

\author{
Eduardo Vera Pichardo\thanks{%
\href{https://orcid.org/0009-0007-7330-3431}{ORCID: 0009-0007-7330-3431}.
B.A.\ in Economics, Universidad Aut\'onoma de Ciudad Ju\'arez (UACJ), Mexico.
Contact: \href{mailto:eduardo546789@gmail.com}{eduardo546789@gmail.com}}\\[2pt]
\normalsize Universidad Aut\'onoma de Ciudad Ju\'arez (UACJ)
}

\date{}

\begin{document}
\maketitle

\begin{abstract}
\noindent Synthetic survey data generated by large language models (LLMs) suffers from a fundamental circularity: the same model family that generates text responses also maps them to numerical scales. We calibrate and validate \textit{Semantic Similarity Rating} (SSR; \citealt{maier2024ssr}), which decouples generation from scale mapping via embedding-based cosine similarity against predefined anchor statements. Configuration experiments ($N = 17$ pilot, $N = 69$ cross-validation across 8 domains) show that naturalistic behavioral anchors outperform formal jargon by 29 percentage points~(pp), and that SSR achieves 65--67\% exact match and 91\% within $\pm$1; a cross-model test with OpenAI \texttt{text-embedding-3-small} reaches 77\% exact, confirming cross-provider generalization. Direct LLM baselines (Claude 87\%, GPT-4o 83\%) establish that SSR's contribution is methodological independence, not accuracy superiority. A control condition removing question text from the LLM prompt actually \textit{improves} LLM accuracy, ruling out information asymmetry as the explanation for SSR's lower accuracy. A pre-registered circularity experiment ($N = 345$) reveals 4-fold compressed error variance in LLM rating ($\sigma^2 = 0.21$ vs.\ $0.87$ for SSR) and systematic directional bias. A cross-model control (GPT-4o rating Claude-generated text) shows nearly identical compression (within/cross ratio $= 0.93$), indicating variance compression is a general LLM property rather than a within-model artifact. The calibration dataset, anchor library, and source code are publicly available (see Data Availability).
\end{abstract}


\newpage

\setlength{\parindent}{0.5in}
\setlength{\emergencystretch}{3em}
\setcounter{secnumdepth}{3}
\setlength{\abovedisplayskip}{18pt}
\setlength{\belowdisplayskip}{18pt}
\setlength{\abovedisplayshortskip}{12pt}
\setlength{\belowdisplayshortskip}{12pt}


\section{Introduction}
\label{sec:introduction}

\subsection{The Promise and Peril of Synthetic Survey Data}

Survey research is the backbone of the behavioral and social sciences, yet it faces mounting challenges: declining response rates \citep{groves2009survey}, rising costs, and increasing difficulty reaching representative samples. Against this backdrop, large language models (LLMs) have emerged as a promising tool for generating synthetic survey responses, simulated data that mimics human respondents' answers to structured questionnaire items \citep{argyle2023out, horton2023llm}.

If LLMs can produce realistic survey data that mirrors the distributional properties of real human samples, researchers could use synthetic data for instrument pretesting, power analysis, hypothesis generation, and, most ambitiously, as a complement to or partial substitute for expensive human data collection \citep{park2023generative}. Several recent studies have demonstrated that LLM-generated responses can reproduce demographic patterns in political attitudes \citep{argyle2023out}, replicate findings from behavioral economics experiments \citep{horton2023llm, aher2023using}, and generate plausible responses to standard survey instruments \citep{bisbee2024synthetic}.

However, this literature has largely overlooked a fundamental methodological problem at the heart of the text-to-scale mapping process: \textit{circularity in measurement}.

\subsection{The Circularity Problem}

The typical pipeline for generating synthetic Likert-scale data proceeds in two steps. First, an LLM generates a natural-language text response to a survey question, given a persona description and contextual instructions. Second, a numeric rating is extracted from this text; this is where the circularity arises. In most implementations, the same model (or same model family) that generated the text also assigns the numeric rating, either through direct instruction (``rate this response on a scale of 1--5'') or through structured output parsing.

This is not independent measurement. It is the model evaluating its own output, analogous to asking a student to grade their own exam. The resulting data conflates two distinct measurement operations that psychometric theory requires to be independent: the \textit{generation} of a response and the \textit{scoring} of that response on a measurement instrument \citep{cronbach1955construct}.

The consequences of this circularity are subtle but consequential. Reliability metrics computed on circularly-validated data will be inflated, because the model's internal consistency reflects its own biases rather than the construct being measured. Validity claims based on agreement between LLM-generated text and LLM-assigned ratings are fundamentally non-transferable, because they measure only the coherence of the model's own outputs rather than alignment with human judgment \citep{maier2024ssr}.

\subsection{Our Contribution}

We systematically calibrate and validate \textbf{Semantic Similarity Rating} (SSR; \citep{maier2024ssr}), a framework that addresses circularity by decoupling text generation from scale mapping, through four key contributions:

 (1)~we implement and validate SSR using Voyage AI \texttt{voyage-3.5-lite} \citep{voyageai2024embeddings} as an architecturally independent measurement channel, confirming cross-provider generalization with OpenAI \texttt{text-embedding-3-small} (77\% vs.\ 67\% exact match); (2)~we demonstrate that naturalistic behavioral anchors outperform formal survey jargon by 29 percentage points (pp); (3)~we provide systematic calibration through factorial experiments identifying the optimal configuration; and (4)~a pre-registered circularity experiment ($N = 345$) with cross-model control reveals 4-fold variance compression in LLM-based rating, established as a general LLM property (within/cross ratio $= 0.93$) rather than a within-model artifact.

On a pilot calibration set ($N = 17$, 3 domains), the final configuration achieves 88\% exact match and 100\% within $\pm 1$ (MAE $= 0.12$). Cross-validation on 69 cases across 8 domains confirms generalization while revealing domain-specific variation (33--90\% exact) that motivates anchor refinement.

\subsection{Paper Overview}

The remainder of this paper is organized as follows. Section~\ref{sec:related} reviews related work on LLMs as survey respondents, text-to-scale mapping approaches, and anchor-based measurement theory. Section~\ref{sec:method} presents the SSR framework architecture and mathematical formulation. Section~\ref{sec:calibration} describes our calibration study, including three systematic configuration experiments, cross-validation analysis, an LLM baseline comparison with information asymmetry control, a cross-model generalization test, and a pre-registered circularity experiment with cross-model control. Section~\ref{sec:discussion} discusses key findings, limitations, and implications. Section~\ref{sec:conclusion} concludes.


\section{Related Work}
\label{sec:related}

\subsection{LLMs as Survey Respondents}

The use of LLMs to simulate survey respondents has rapidly expanded since the seminal work of \citet{argyle2023out}, who demonstrated that GPT-3 could produce response distributions conditioned on demographic variables that closely matched those of real American subpopulations across political, economic, and social attitude items. Subsequent work has extended this approach to economic decision-making \citep{horton2023llm}, social behavior simulation \citep{park2023generative, aher2023using}, and political survey replication \citep{bisbee2024synthetic}.

However, critical evaluations have revealed important limitations. \citet{santurkar2023whose} showed that LLM-generated opinion distributions are systematically biased toward certain demographic groups, reflecting the composition and biases of training data rather than any genuine model of human attitudes. \citet{bisbee2024synthetic} demonstrated that synthetic survey replacements can produce misleading results on novel political questions, particularly when the question format or content departs from the model's training distribution.

A critical gap in this literature is the relative inattention to the \textit{measurement pipeline}, specifically how text responses are mapped to numeric scale values. Most studies either instruct the LLM to produce numeric output directly, parse structured responses, or use the same (or similar) model to rate its own output. The SSR framework proposed by \citet{maier2024ssr} represents the closest precursor to our work. They demonstrated that embedding-based semantic similarity can map LLM-generated text to Likert distributions that closely replicate real human survey data, testing on 57 personal care product surveys with 9,300 human responses as ground truth from a leading consumer goods corporation. Their validation is \textit{distributional}: they show that synthetic response distributions match human distributions at the population level (Kolmogorov--Smirnov similarity $> 0.85$, achieving 90\% of human test--retest reliability). This is a powerful result that establishes SSR's viability for applied market research, but it addresses a different question than individual-level accuracy: distributional similarity can hold even when individual predictions contain errors, because overestimates and underestimates cancel at the aggregate level.

Our work complements theirs by addressing the \textit{engineering} and \textit{methodological} questions that distributional validation leaves open. First, we provide systematic calibration through factorial experiments that identify \textit{why} SSR works, discovering that anchor statement quality (+29~pp) and asymmetric embedding (+6~pp) are the dominant factors, rather than demonstrating \textit{that} it works at the distributional level. Second, we frame the contribution explicitly in terms of the circularity problem: the measurement independence that embedding-based mapping provides relative to LLM self-rating. Notably, \citeauthor{maier2024ssr}'s comparison of follow-up Likert rating (where the same model generates and rates text) against SSR shows inferior distributional similarity for the self-rating approach, a result consistent with our finding that self-rating circularity introduces measurable bias, but they do not identify circularity as the mechanism driving this difference. Third, we generalize beyond their purchase intent focus to 8 semantic domains, revealing substantial domain-specific variation (33--90\% exact match) that motivates targeted anchor refinement. Finally, we compare SSR against a direct LLM baseline with controlled methodology (temperature $= 0$, two prompt variants), establishing the accuracy--independence tradeoff that practitioners must navigate.

\subsection{Text Embeddings and Semantic Similarity}

Text embeddings, dense vector representations of natural language that encode semantic content in continuous space, provide the mathematical foundation for SSR. Modern embedding models produce vectors where cosine similarity correlates with semantic relatedness \citep{reimers2019sentence}.

The quality of these representations has improved substantially with advances in contrastive learning. \citet{gao2021simcse} introduced SimCSE, demonstrating that simple contrastive objectives could substantially improve sentence embedding quality. The Massive Text Embedding Benchmark \citep{muennighoff2023mteb} established standardized evaluation across retrieval, classification, and semantic similarity tasks, enabling systematic comparison of embedding models.

A critical property of text embeddings for our application is \textit{compression}: semantically related texts, even those expressing opposite sentiments within the same domain, produce cosine similarities clustered in a narrow range (typically spanning only 0.05--0.10 units for symmetric embedding). This compression is a known property of high-dimensional embedding spaces and requires normalization strategies to enable meaningful discrimination between scale points, as we demonstrate in Section~\ref{sec:calibration}.

Modern embedding models also support \textit{asymmetric embedding}, where query and document texts are embedded with different objectives \citep{wang2022text}. This is standard practice in information retrieval, where queries and documents have different distributional properties. We exploit this capability by embedding response texts as queries and anchor statements as documents, treating scale mapping as a retrieval problem.

\subsection{Anchor-Based Measurement in Psychology}

The design of anchor statements for Likert scales has a long history in psychometrics \citep{likert1932technique}. \citet{krosnick2010question} established that the specific language of scale endpoints and anchors significantly affects response behavior, with behavioral descriptions producing more consistent and valid responses than abstract labels.

\citet{tourangeau2000psychology} proposed a cognitive model of survey response in which respondents (1) comprehend the question, (2) retrieve relevant information, (3) form a judgment, and (4) map that judgment to the response scale. The fourth step, response mapping, is where anchor language exerts its influence. Clear, concrete anchors reduce mapping ambiguity and improve measurement precision.

This literature is directly relevant to SSR, although the mechanism is different. In human survey response, anchor language affects cognitive mapping from judgment to scale point. In SSR, anchor language affects \textit{embedding distance} between response text and scale point representations. Our finding that naturalistic behavioral anchors substantially outperform formal survey jargon (Section~\ref{sec:anchor-quality}) can be understood as the embedding-space analog of the cognitive anchoring effect: richer, more distinctive language provides better ``targets'' for similarity-based mapping, just as it provides clearer cognitive reference points for human respondents.


\section{Method: Semantic Similarity Rating}
\label{sec:method}

\subsection{Architecture Overview}

\begin{figure}[htbp]
\centering
\includegraphics[width=\textwidth,trim=8 8 8 8,clip]{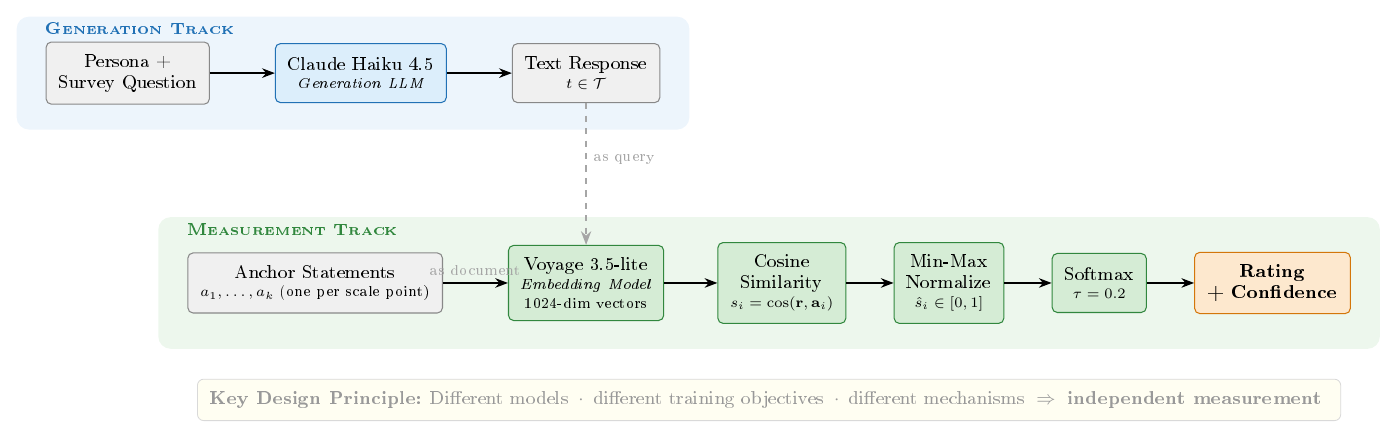}
\caption{SSR pipeline architecture. The generation model (Claude Haiku 4.5) produces text responses; the embedding model independently maps them to scale ratings via cosine similarity. The primary configuration uses Voyage 3.5-lite (1024 dims); cross-model validation with OpenAI text-embedding-3-small (1536 dims) is reported in Section~\ref{sec:cross-model}. The two models are architecturally independent, substantially reducing circular validation.}
\label{fig:architecture}
\end{figure}

The SSR framework implements a two-model pipeline that separates text generation from scale measurement. Figure~\ref{fig:architecture} illustrates the full architecture.

\begin{enumerate}
\item \textbf{Generation}: A large language model (Claude Haiku 4.5) generates a natural-language text response given a persona description and survey question. The generation prompt includes scale context (anchors for the lowest and highest points) to align the text's intensity with the intended scale semantics.

\item \textbf{Measurement}: An embedding model (Voyage AI \texttt{voyage-3.5-lite}, 1024 dimensions) independently maps the generated text to a numeric rating through cosine similarity against predefined anchor statements.
\end{enumerate}

\needspace{4\baselineskip}
The key property is that the generation and measurement models are architecturally independent: they use different model weights, different training objectives, and fundamentally different computational mechanisms (autoregressive text generation vs. contrastive embedding). This substantially reduces the circular dependency present in single-model approaches, though we note that shared pretraining corpora between embedding and generation models may introduce residual correlations (see Section~\ref{sec:discussion}).

\subsection{Scale Anchor System}

The anchor system provides natural-language statements representing each point on a rating scale. For a $k$-point scale, we define $k$ anchor statements $\{a_1, a_2, \ldots, a_k\}$ ordered from the lowest to highest construct level.

\subsubsection{Anchor Families}

We define 15 \textit{semantic families} covering the most common constructs in survey research: satisfaction, likelihood, agreement, ease, importance, familiarity, appeal, value, trust, quality, frequency, uniqueness, relevance, impression, and purchase intent, plus an 11-point NPS (Net Promoter Score) family (included in the framework but not evaluated in this study due to its different scale structure).

Each family contains five anchor statements (one per scale point for 5-point Likert items) written in first-person naturalistic behavioral language. For example, the satisfaction family ranges from:

\begin{quote}
\textit{``Terrible. I was frustrated and angry the entire time, nothing worked right and I regret using this.''} (rating = 1)
\end{quote}

to:

\begin{quote}
\textit{``Excellent. Everything was smooth, fast, and easy. I'm really impressed and delighted with the result.''} (rating = 5)
\end{quote}

\subsubsection{Resolution Hierarchy}

Anchor resolution follows a deterministic fallback chain:
\begin{enumerate}
\item \textbf{Predefined templates}: The question's semantic family is identified from its type or label anchors, and the corresponding hand-written template is used.
\item \textbf{Label-based lookup}: If the question specifies custom scale anchor labels (e.g., ``Not at all'' to ``Completely''), these labels are matched to a semantic family via a lookup table.
\item \textbf{Interpolation fallback}: For fully custom scales with no template match, anchors are generated by interpolating between the provided low and high labels.
\end{enumerate}

\subsection{Embedding and Similarity Computation}

\subsubsection{Embedding Model}

We use Voyage AI's \texttt{voyage-3.5-lite} model as the primary embedding provider, which produces 1024-dimensional vectors trained for semantic similarity and information retrieval tasks. We selected this model for three reasons: (1) it ranks competitively on the Massive Text Embedding Benchmark (MTEB; \citealp{muennighoff2023mteb}) across retrieval and semantic similarity tasks; (2) it natively supports asymmetric embedding via an \texttt{input\_type} parameter that optimizes vectors differently for \texttt{document} and \texttt{query} roles, which we exploit for scale mapping; and (3) it offers favorable cost characteristics (\$0.02 per million tokens) suitable for production deployment. To verify that SSR's effectiveness is not specific to Voyage, we also evaluate with OpenAI's \texttt{text-embedding-3-small} (1536 dimensions), which does not support asymmetric embedding, providing a fair symmetric-to-symmetric comparison (Section~\ref{sec:cross-model}).

\subsubsection{Asymmetric Embedding}

Anchor statements are embedded as \texttt{document} type, and response texts are embedded as \texttt{query} type. This treats scale mapping as a retrieval task: given a response (query), retrieve the most similar anchor (document). Empirically, asymmetric embedding widens the similarity spread from $\sim$0.05 to $\sim$0.15, improving discrimination between scale points (Section~\ref{sec:ablation}).

\subsubsection{Cosine Similarity}

For a response vector $\mathbf{r}$ and anchor vectors $\mathbf{a}_1, \ldots, \mathbf{a}_k$, we compute:

\begin{equation}
s_i = \frac{\mathbf{r} \cdot \mathbf{a}_i}{\|\mathbf{r}\| \cdot \|\mathbf{a}_i\|}
\label{eq:cosine}
\end{equation}

yielding a similarity vector $\mathbf{s} = (s_1, \ldots, s_k)$.

\subsection{From Similarities to Distributions}

\subsubsection{The Compression Problem}

\begin{figure}[htbp]
\centering
\includegraphics[width=0.95\textwidth]{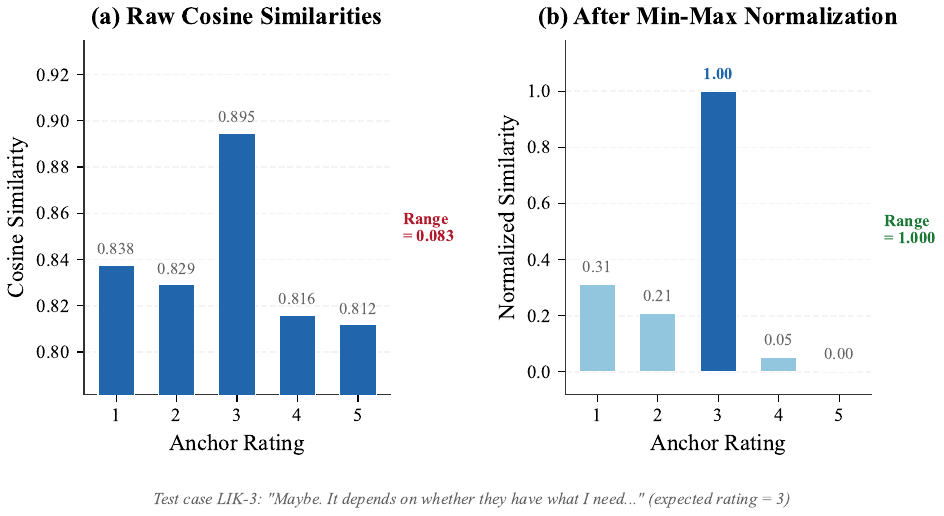}
\caption{Cosine similarity compression illustrated on test case LIK-3 (expected rating = 3). (a)~Raw cosine similarities between the response text and all 5 anchor statements span only 0.083 (from 0.812 to 0.895), making the correct anchor difficult to identify. (b)~After min-max normalization, the same similarities are stretched to $[0, 1]$: the correct anchor (Rating~3) reaches 1.00, clearly separated from the next-highest (Rating~1, 0.31).}
\label{fig:compression}
\end{figure}

Raw cosine similarities between text embeddings cluster in a narrow range, typically spanning only 0.05--0.10 units (e.g., $[0.75, 0.83]$ or $[0.81, 0.90]$ depending on the specific texts and embedding model). This \textit{compression} phenomenon, a known property of high-dimensional embedding spaces, means that a na\"ive softmax over raw similarities produces near-uniform distributions that lack discriminative power. Figure~\ref{fig:compression} illustrates this problem on a representative test case.

\subsubsection{Min-Max Normalization}

We address compression by applying min-max normalization to stretch the similarity range to $[0, 1]$:

\begin{equation}
\hat{s}_i = \frac{s_i - \min(\mathbf{s})}{\max(\mathbf{s}) - \min(\mathbf{s})}
\label{eq:min-max}
\end{equation}

This preserves the ordinal relationships between similarities while amplifying relative differences.

\subsubsection{Softmax with Temperature}

Normalized similarities are converted to a probability distribution using softmax with temperature $\tau$:

\begin{equation}
p_i = \frac{\exp(\hat{s}_i / \tau)}{\sum_{j=1}^{k} \exp(\hat{s}_j / \tau)}
\label{eq:softmax}
\end{equation}

Lower $\tau$ produces more peaked (decisive) distributions; higher $\tau$ produces more uniform (uncertain) ones. The pilot study selected $\tau = 0.2$ for its favorable within $\pm$1 performance (Section~\ref{sec:exp1}); subsequent cross-validation on 8 domains identified $\tau = 0.15$ as optimal (Section~\ref{sec:cross-validation}).

\subsubsection{Rating Extraction}

The final rating is computed as the weighted mean of scale points, rounded to the nearest integer:

\begin{equation}
\text{rating} = \text{round}\left(\sum_{i=1}^{k} p_i \cdot (i + r_{\min} - 1)\right)
\label{eq:rating}
\end{equation}

\noindent where $r_{\min}$ is the minimum scale value (typically 1 for Likert items). We use the weighted mean (expected value) rather than argmax because it produces more calibrated predictions for ambiguous cases where probability mass is distributed across adjacent scale points. An argmax approach yields identical results when the softmax distribution is peaked (high confidence) but loses the interpolation that helps SSR handle boundary cases. The two approaches diverge most for intermediate-confidence predictions; a systematic comparison is left for future work.

\subsubsection{Confidence Estimation}

We estimate confidence using normalized Shannon entropy:

\begin{equation}
c = 1 - \frac{H(\mathbf{p})}{H_{\max}} = 1 - \frac{-\sum_{i} p_i \log_2 p_i}{\log_2 k}
\label{eq:confidence}
\end{equation}

\noindent A fully peaked distribution ($p_i = 1$ for exactly one $i$) yields $c = 1$; a uniform distribution yields $c = 0$. This provides a calibrated measure of mapping certainty.

\subsection{Methodology Versioning}

Each generated response is tagged with complete methodology metadata: generation model, mapping method (\texttt{embedding-v1} for SSR), embedding model identifier, temperature, normalization method, and a methodology version string. This enables factorial comparisons across configurations and ensures full reproducibility.

\subsection{Model and API Details}

Table~\ref{tab:model-details} lists the specific models and API parameters used throughout this study. Text generation and LLM rating used Claude Haiku 4.5 \citep{anthropic2024claude3}; cross-model validation used GPT-4o \citep{openai2023gpt4}; embeddings used Voyage 3.5-lite \citep{voyageai2024embeddings} and OpenAI text-embedding-3-small. All API calls were made between January and February 2026.

\begin{table}[htbp]
\centering
\caption{Model and API details. All experiments used deterministic settings where available.}
\label{tab:model-details}
\scriptsize
\begin{tabular}{llll}
\toprule
Role & Model / Endpoint & Provider & Key Parameters \\
\midrule
Text generation & \texttt{claude-haiku-4-5-20251001} & Anthropic & $T = 0.7$, max\_tokens $= 500$ \\
LLM rating & \texttt{claude-haiku-4-5-20251001} & Anthropic & $T = 0$, max\_tokens $= 100$ \\
LLM rating & \texttt{gpt-4o}\textsuperscript{a} & OpenAI & $T = 0$, max\_tokens $= 100$ \\
Embedding (primary) & \texttt{voyage-3.5-lite} (1024d) & Voyage AI & input\_type: query/document \\
Embedding (validation) & \texttt{text-embedding-3-small} (1536d) & OpenAI & --- \\
\bottomrule
\multicolumn{4}{l}{\scriptsize \textsuperscript{a} The \texttt{gpt-4o} alias resolves to OpenAI's latest snapshot; experiments conducted Jan--Feb 2026.}
\end{tabular}
\normalsize
\end{table}


\section{Calibration Study}
\label{sec:calibration}

We conducted a systematic calibration study in two phases. First, a pilot study with 17 test cases across 3 domains guided configuration search through three factorial experiments. Second, the winning configuration was evaluated on an expanded 69-case test set across 8 domains, which provides the primary evaluation of SSR's accuracy and generalization.

\subsection{Test Set Construction}

\subsubsection{Test Cases}

We constructed two test sets. The \textit{pilot set} comprises 17 test cases across three semantic domains: satisfaction (7 cases), likelihood (5 cases), and agreement (5 cases). Each test case consists of a survey question, a human-written response text, and an expected rating on a 1--5 Likert scale. This set served as the development sample for configuration search (Experiments 1--3 below).

Test cases were designed to cover:

\begin{itemize}
\item \textbf{Clear polarity}: Extreme ratings (1 and 5) with unambiguous sentiment signals.
\item \textbf{Neutral ambiguity}: Midpoint ratings (3) with hedging and mixed language.
\item \textbf{Subtle mixed signals}: Adjacent ratings (2 and 4) with hedging, qualified praise, or mild criticism that makes the ``correct'' rating less obvious.
\end{itemize}

Representative examples from each domain and difficulty level are shown in Table~\ref{tab:test-cases}.

\begin{table}[htbp]
\centering
\caption{Representative reference test cases. Each case has a human-assigned expected rating and difficulty level (clear/subtle).}
\label{tab:test-cases}
\begin{tabular}{lp{7cm}cc}
\toprule
Domain & Response Text (truncated) & Expected & Difficulty \\
\midrule
Satisfaction & ``Absolutely awful. The page froze three times...'' & 1 & Clear \\
Satisfaction & ``I managed to complete my order but it was frustrating...'' & 2 & Subtle \\
Satisfaction & ``It was okay. Nothing terrible but nothing great...'' & 3 & Clear \\
Likelihood & ``Maybe. It depends on whether they have what I need...'' & 3 & Clear \\
Likelihood & ``Yeah, I'll probably order again. They had decent prices...'' & 4 & Clear \\
Agreement & ``I don't really agree. The quality is below what I expected...'' & 2 & Clear \\
Agreement & ``I mostly agree. Well-made and reliable. A few minor things...'' & 4 & Clear \\
\bottomrule
\end{tabular}
\end{table}

\subsubsection{Expanded Test Set}

To assess generalization beyond the initial 3-domain calibration, we constructed an expanded test set of 69 cases across 8 semantic domains (Table~\ref{tab:expanded-testset}). The additional domains (ease, importance, trust, value, and purchase intent) were selected to cover the most common Likert constructs in market research and behavioral science. Each domain includes clear (unambiguous polarity), subtle (mixed signals near adjacent scale points), and edge (extreme cases or implicit sentiment) test cases. All domains include clear cases for every rating 1--5 to ensure balanced coverage. Seven additional clear-difficulty cases were added to the original three domains to ensure that every domain covers all five rating levels, bringing the total from 62 to 69.

\begin{table}[htbp]
\centering
\caption{Expanded test set composition: 69 cases across 8 semantic domains with difficulty stratification.}
\label{tab:expanded-testset}
\begin{tabular}{lccccc}
\toprule
Domain & $N$ & Clear & Subtle & Edge & Rating Distribution \\
\midrule
Satisfaction & 10 & 5 & 3 & 2 & 1--5 balanced \\
Likelihood & 8 & 5 & 2 & 1 & 1--5 balanced \\
Agreement & 8 & 5 & 2 & 1 & 1--5 balanced \\
Ease & 9 & 5 & 2 & 2 & 1--5 balanced \\
Importance & 9 & 5 & 2 & 2 & 1--5 balanced \\
Trust & 9 & 5 & 2 & 2 & 1--5 balanced \\
Value & 8 & 5 & 2 & 1 & 1--5 balanced \\
Purchase intent & 8 & 5 & 2 & 1 & 1--5 balanced \\
\midrule
\textbf{Total} & \textbf{69} & \textbf{40} & \textbf{17} & \textbf{12} & \\
\bottomrule
\end{tabular}
\end{table}

The 17-case pilot set was sufficient for comparative evaluation across configurations (Experiments 1--3) but is too small for generalization claims. The expanded 69-case set serves as the \textit{primary evaluation} of SSR accuracy and is used for cross-validation analysis (Section~\ref{sec:cross-validation}) and baseline comparison (Section~\ref{sec:baseline}).

\subsection{Experiment 1: Normalization and Temperature}
\label{sec:exp1}

\subsubsection{Design}

We tested 13 configurations crossing 3 normalization methods (none, min-max, z-score) with multiple temperature values:

\begin{itemize}
\item \textbf{No normalization}: Raw cosine similarities, $\tau \in \{0.01, 0.005, 0.001\}$
\item \textbf{Min-max normalization}: Eq.~\ref{eq:min-max}, $\tau \in \{1.0, 0.5, 0.2, 0.1, 0.05, 0.01\}$
\item \textbf{Z-score normalization}: Center and scale, $\tau \in \{1.0, 0.5, 0.2, 0.1\}$
\end{itemize}

All configurations used the same anchor statements (formal jargon, the default at the time of this experiment) and symmetric embedding (\texttt{document/document}).

\subsubsection{Results}

Of the 13 configurations tested, Table~\ref{tab:grid-search} reports the top six sorted by exact match accuracy.

\begin{table}[htbp]
\centering
\caption{Grid search over normalization and temperature. All configurations used formal anchor statements and symmetric embedding. $N = 17$ test cases.}
\label{tab:grid-search}
\begin{tabular}{llccc}
\toprule
Normalization & $\tau$ & Exact & Within $\pm$1 & MAE \\
\midrule
none & 0.005 & 9/17 (53\%) & 15/17 (88\%) & 0.53 \\
min-max & 0.05 & 9/17 (53\%) & 15/17 (88\%) & 0.53 \\
min-max & 0.1 & 8/17 (47\%) & 16/17 (94\%) & 0.59 \\
zscore & 0.2 & 8/17 (47\%) & 16/17 (94\%) & 0.59 \\
min-max & 0.2 & 7/17 (41\%) & 17/17 (100\%) & 0.59 \\
none & 0.001 & 7/17 (41\%) & 14/17 (82\%) & 0.71 \\
\bottomrule
\end{tabular}
\end{table}

\subsubsection{Findings}

Two configurations tied at 53\% exact match: \texttt{none}/$\tau=0.005$ and \texttt{min-max}/$\tau=0.05$. Min-max normalization produced more stable results across temperature values, with 100\% within $\pm$1 at $\tau = 0.2$. We selected \texttt{min-max}/$\tau=0.2$ as the default due to its favorable within $\pm$1 rate and interpretable confidence scores.

The 53\% accuracy ceiling with formal anchors motivated our investigation of anchor statement quality (Experiment~2).

\subsection{Experiment 2: Anchor Statement Quality}
\label{sec:anchor-quality}

\subsubsection{Hypothesis}

We hypothesized that the formal, abstract language typical of survey anchor labels (e.g., ``I am very dissatisfied. The experience was poor.'') would produce poor embedding discrimination because such statements are semantically similar to each other. In contrast, naturalistic behavioral descriptions (e.g., ``Terrible. I was frustrated and angry the entire time, nothing worked right and I regret using this.'') should produce more distinctive embeddings, improving similarity spread and discrimination.

\subsubsection{Manipulation}

We rewrote anchor statements for three semantic families (satisfaction, likelihood, agreement) from formal survey jargon to first-person naturalistic behavioral language. The rewriting followed three principles:

\begin{itemize}
\item \textbf{Specificity}: Replace abstract labels with concrete behavioral descriptions.
\item \textbf{Emotion}: Include affective language that captures the feeling, not just the evaluation.
\item \textbf{Distinctiveness}: Ensure that adjacent scale points (e.g., ratings 2 and 3) produce maximally different embeddings while remaining semantically appropriate.
\end{itemize}

Table~\ref{tab:anchor-transform} illustrates the transformation for the satisfaction family. Formal anchors share a common syntactic template (``I am [adverb] [dis]satisfied''), producing nearly identical embeddings (mean inter-anchor cosine similarity $\approx 0.89$). Naturalistic anchors use varied vocabulary and emotional language, reducing mean inter-anchor similarity to $\approx 0.82$ and increasing discrimination.

\begin{table}[htbp]
\centering
\caption{Formal vs.\ naturalistic anchor statements for the satisfaction family. Naturalistic anchors use varied vocabulary, concrete behavioral descriptions, and emotional language to maximize embedding discrimination.}
\label{tab:anchor-transform}
\begin{tabular}{cp{5.5cm}p{5.5cm}}
\toprule
Rating & Formal Anchor & Naturalistic Anchor \\
\midrule
1 & ``I am very dissatisfied. The experience was poor.'' & ``Terrible. I was frustrated and angry the entire time, nothing worked right and I regret using this.'' \\
2 & ``I am somewhat dissatisfied. The experience was below average.'' & ``Not great. I ran into problems and it was harder than it should have been, below what I expected.'' \\
3 & ``I am neither satisfied nor dissatisfied. The experience was average.'' & ``It was okay. Nothing special, not bad but not good either. Pretty average and forgettable.'' \\
4 & ``I am somewhat satisfied. The experience was above average.'' & ``Pretty good. Things went smoothly for the most part, just a couple of small things that could be better.'' \\
5 & ``I am very satisfied. The experience was excellent.'' & ``Excellent. Everything was smooth, fast, and easy. I'm really impressed and delighted with the result.'' \\
\bottomrule
\end{tabular}
\end{table}

\subsubsection{Results}

\begin{figure}[htbp]
\centering
\includegraphics[width=0.95\textwidth]{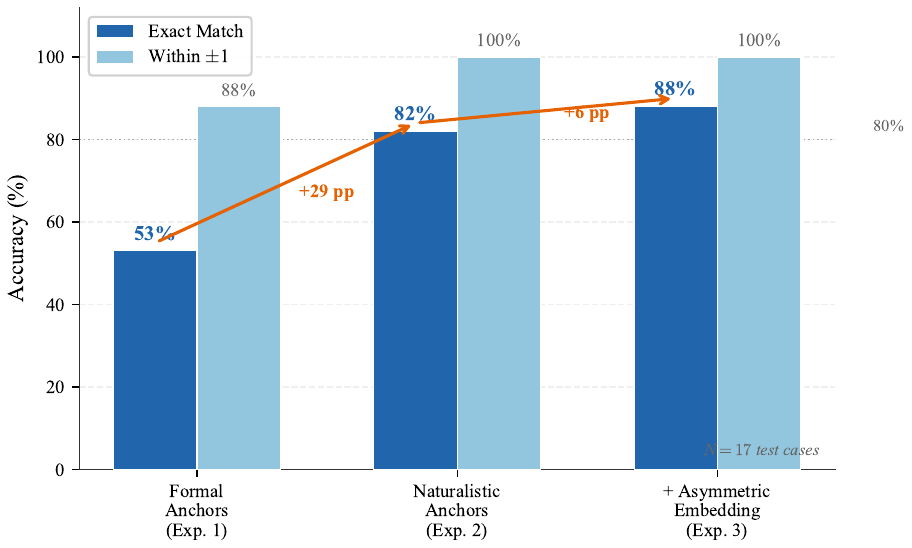}
\caption{Accuracy progression across the three calibration experiments. Naturalistic anchors account for the largest improvement (+29~pp), followed by asymmetric embedding (+6~pp). Light bars indicate within $\pm$1 accuracy.}
\label{fig:accuracy}
\end{figure}

As shown in Table~\ref{tab:anchor-quality}, naturalistic anchors substantially outperform formal anchors; Figure~\ref{fig:accuracy} traces the cumulative accuracy gains across all three calibration experiments.

\begin{table}[htbp]
\centering
\caption{Effect of anchor statement quality. Naturalistic behavioral anchors outperform formal survey jargon across all metrics. Best configuration for each anchor type is shown.}
\label{tab:anchor-quality}
\begin{tabular}{llcccc}
\toprule
Anchors & Config & Exact & Within $\pm$1 & MAE & Sim.\ Spread \\
\midrule
Formal & none/$\tau$=0.005 & 9/17 (53\%) & 15/17 (88\%) & 0.53 & 0.05 \\
Naturalistic & min-max/$\tau$=0.2 & 14/17 (82\%) & 17/17 (100\%) & 0.18 & 0.09 \\
\midrule
\multicolumn{2}{l}{Improvement} & +29~pp & +12~pp & $-$0.35 & +80\% \\
\bottomrule
\end{tabular}
\end{table}

\subsubsection{Mechanism}

The improvement is explained by increased similarity spread. Formal anchors produce cosine similarities clustered in a $\sim$0.05 range (e.g., [0.78, 0.83]), providing little signal for discrimination. Naturalistic anchors widen this to $\sim$0.09 (e.g., [0.74, 0.83]), enabling the softmax to produce more peaked distributions that correctly identify the matching scale point.

\subsubsection{Implication}

\textbf{Anchor quality is the single most impactful factor in SSR accuracy}, more impactful than normalization method, temperature tuning, or embedding strategy. This finding has direct implications for any embedding-based approach to text classification against semantic prototypes.

\vspace{-0.5\baselineskip}
\subsection{Experiment 3: Ablation Study}
\label{sec:ablation}

\subsubsection{Design}

We conducted a $2 \times 2$ factorial experiment crossing two proposed optimizations:

\begin{itemize}
\item \textbf{H2 (Contextualized anchors)}: Prepend the survey question text to each anchor statement before embedding (e.g., ``Question: How satisfied are you? Answer: Terrible...'').
\item \textbf{H3 (Asymmetric embedding)}: Embed anchor statements as \texttt{document} and response texts as \texttt{query}, rather than both as \texttt{document}.
\end{itemize}

Both optimizations used naturalistic anchors and \texttt{min-max}/$\tau=0.2$.

\subsubsection{Results}

The $2 \times 2$ factorial results are summarized in Table~\ref{tab:ablation}.

\begin{table}[htbp]
\centering
\caption{$2 \times 2$ factorial ablation. H2: question-contextualized anchors. H3: asymmetric embedding (anchors as document, responses as query). Bold indicates best condition.}
\label{tab:ablation}
\begin{tabular}{llcccc}
\toprule
Condition & Context & Asymmetric & Exact & Within $\pm$1 & MAE \\
\midrule
A: baseline & No & No & 14/17 (82\%) & 17/17 (100\%) & 0.18 \\
B: H2 only & Yes & No & 11/17 (65\%) & 17/17 (100\%) & 0.35 \\
\textbf{C: H3 only} & \textbf{No} & \textbf{Yes} & \textbf{15/17 (88\%)} & \textbf{17/17 (100\%)} & \textbf{0.12} \\
D: H2+H3 & Yes & Yes & 12/17 (71\%) & 17/17 (100\%) & 0.29 \\
\bottomrule
\end{tabular}
\end{table}

\subsubsection{Findings}

\begin{figure}[htbp]
\centering
\includegraphics[width=0.95\textwidth]{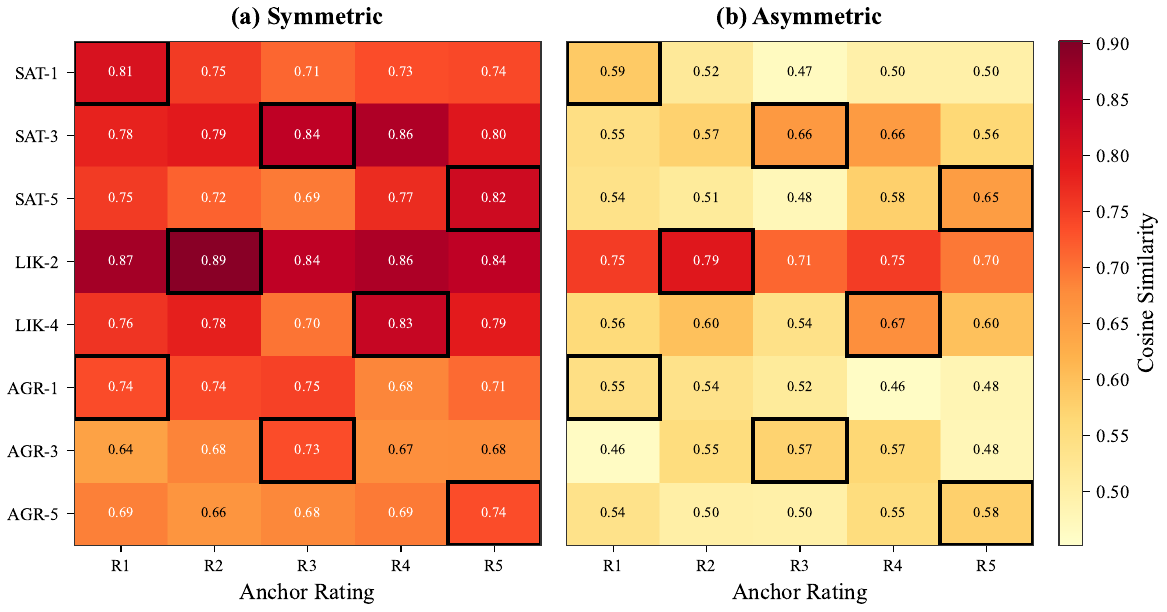}
\caption{Similarity heatmaps for symmetric (left) vs.\ asymmetric (right) embedding. Each cell shows the cosine similarity between a test response and the 5 anchor statements. Asymmetric embedding produces wider similarity spread, improving discrimination between adjacent scale points.}
\label{fig:heatmap}
\end{figure}

\begin{itemize}
\item \textbf{H3 (asymmetric) improves accuracy}: Condition C achieves 88\% exact match, a 6 percentage point improvement over the baseline (82\%). The improvement comes from wider similarity spread in asymmetric mode, where the query embedding objective optimizes for discriminative retrieval. Figure~\ref{fig:heatmap} visualizes this contrast: symmetric embedding produces compressed similarity ranges, while asymmetric embedding widens them.

\item \textbf{H2 (context) is harmful}: Adding question text to anchor statements reduces accuracy by 17 percentage points (82\% $\to$ 65\%). The question text introduces shared semantic content that makes all anchors more similar to each other, reducing the discriminability between scale points. Alternative contextualization strategies (such as embedding question and response jointly while keeping anchors context-free, or using learned question-conditioned anchor weighting) were not explored and may yield different results.

\item \textbf{Interaction}: H2's negative effect partially cancels H3's positive effect in Condition D (71\%), confirming that na\"ive contextualization should not be combined with asymmetric embedding.
\end{itemize}

\subsection{Final Configuration (Pilot Result)}

Based on the three pilot experiments ($N = 17$), we established the following optimal SSR configuration:

\begin{itemize}
\item \textbf{Generation model}: Claude Haiku 4.5
\item \textbf{Embedding model}: Voyage AI \texttt{voyage-3.5-lite} (1024 dims)
\item \textbf{Embedding strategy}: Asymmetric (anchors = document, responses = query)
\item \textbf{Normalization}: Min-max $\to [0, 1]$
\item \textbf{Temperature}: $\tau = 0.2$ (pilot default; subsequent cross-validation on 8 domains identified $\tau = 0.15$ as optimal; see Section~\ref{sec:cross-validation}. We report results at both values throughout for transparency)
\item \textbf{Anchor language}: Naturalistic first-person behavioral statements
\end{itemize}

\noindent \textbf{Pilot result}: 15/17 (88\%) exact match, 17/17 (100\%) within $\pm$1, MAE $= 0.12$. We emphasize that this is a configuration-search result on the development sample; the primary evaluation on the held-out 69-case set follows below.

The two pilot errors occurred on subtle cases (mixed signals) where the model predicted one scale point away from the expected rating. In both cases, the predicted rating was arguably defensible given the ambiguity of the response text.

\subsection{Cross-Validation on Expanded Test Set}
\label{sec:cross-validation}

To evaluate generalization beyond the 3 domains used for configuration tuning, we applied the final SSR configuration (asymmetric embedding, naturalistic anchors, min-max normalization) to the expanded 69-case test set across 8 domains. We conducted two complementary validation analyses.

\subsubsection{Global Performance}

Accuracy across all 69 cases at multiple temperature values is reported in Table~\ref{tab:cv-global}.

\begin{table}[htbp]
\centering
\caption{Global accuracy on expanded 69-case test set. Asymmetric embedding with naturalistic anchors and min-max normalization. Bold indicates best exact match.}
\label{tab:cv-global}
\begin{tabular}{lccc}
\toprule
$\tau$ & Exact & Within $\pm$1 & MAE \\
\midrule
0.05 & 45/69 (65\%) & 62/69 (90\%) & 0.46 \\
0.10 & 45/69 (65\%) & 62/69 (90\%) & 0.45 \\
\textbf{0.15} & \textbf{46/69 (67\%)} & \textbf{63/69 (91\%)} & \textbf{0.42} \\
0.20 & 45/69 (65\%) & 63/69 (91\%) & 0.43 \\
0.25 & 43/69 (62\%) & 63/69 (91\%) & 0.46 \\
0.30 & 39/69 (57\%) & 63/69 (91\%) & 0.52 \\
\bottomrule
\end{tabular}
\end{table}

Overall accuracy on the expanded set is substantially lower than the initial 3-domain result (67\% vs.\ 88\% exact), reflecting the inclusion of harder domains and more diverse construct types. However, within $\pm$1 accuracy remains high at 91\%, indicating that the framework reliably identifies the correct region of the scale.

\subsubsection{Per-Domain Analysis}

Per-domain results reveal substantial variation (Table~\ref{tab:cv-domain}).

\begin{table}[htbp]
\centering
\caption{Per-domain accuracy at $\tau = 0.2$ (pilot default). Cross-validation subsequently identified $\tau = 0.15$ as optimal (Table~\ref{tab:cv-global}); per-domain patterns are substantively identical at both values. Sorted by descending exact match accuracy; the original 3 calibration domains (satisfaction, likelihood, agreement) are interspersed with the 5 new domains. Note: With $N = 8$--10 per domain, individual domain estimates have wide binomial confidence intervals (e.g., 95\% CI for 90\% exact with $N = 10$ is [56\%, 100\%]). Per-domain results should be interpreted as indicative of tier patterns rather than precise point estimates.}
\label{tab:cv-domain}
\begin{tabular}{lcccc}
\toprule
Domain & $N$ & Exact & Within $\pm$1 & MAE \\
\midrule
Satisfaction & 10 & 9/10 (90\%) & 10/10 (100\%) & 0.10 \\
Purchase intent & 8 & 7/8 (88\%) & 8/8 (100\%) & 0.13 \\
Agreement & 8 & 6/8 (75\%) & 8/8 (100\%) & 0.25 \\
Value & 8 & 6/8 (75\%) & 8/8 (100\%) & 0.25 \\
Likelihood & 8 & 5/8 (63\%) & 8/8 (100\%) & 0.38 \\
Ease & 9 & 5/9 (56\%) & 6/9 (67\%) & 0.78 \\
Trust & 9 & 4/9 (44\%) & 9/9 (100\%) & 0.56 \\
Importance & 9 & 3/9 (33\%) & 6/9 (67\%) & 1.00 \\
\bottomrule
\end{tabular}
\end{table}

\begin{figure}[htbp]
\centering
\includegraphics[width=0.95\textwidth]{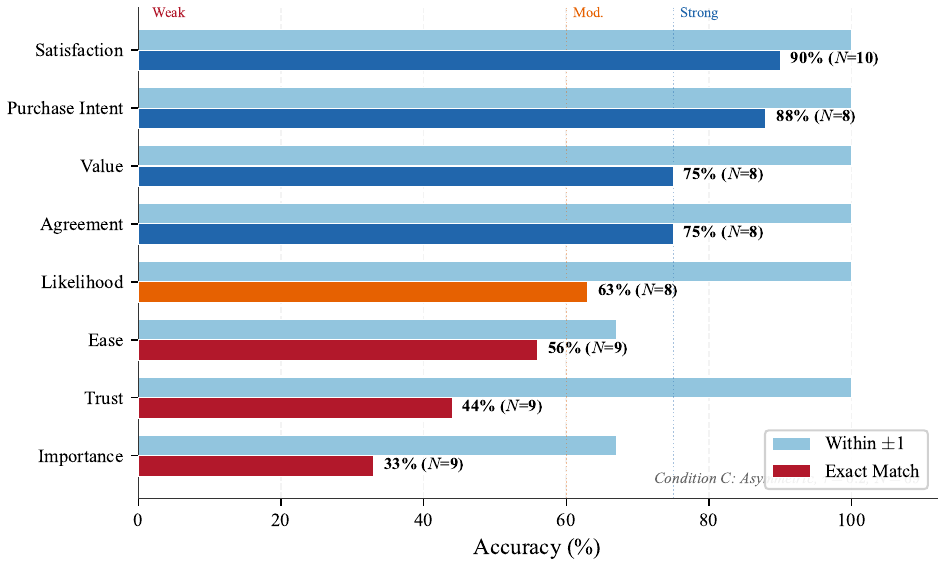}
\caption{Per-domain exact match accuracy on the expanded 69-case test set. Three performance tiers are visible: strong (satisfaction, purchase intent), moderate (agreement, value, likelihood), and weak (ease, trust, importance).}
\label{fig:per-domain}
\end{figure}

Three performance tiers are evident (Figure~\ref{fig:per-domain}): (1) \textit{strong} domains (satisfaction, purchase intent; 88--90\% exact) where naturalistic anchors provide clear discrimination; (2) \textit{moderate} domains (agreement, value, likelihood; 63--75\% exact, 100\% $\pm$1) where the framework captures the correct scale region but sometimes selects adjacent points; and (3) \textit{weak} domains (ease, trust, importance; 33--56\% exact) where the anchor statements may require further refinement.

By difficulty level: clear cases achieve 70\% exact match, subtle cases 59\%, and edge cases 58\%, confirming that the framework degrades gracefully with ambiguity.

\paragraph{Qualitative error patterns.}
SSR's errors concentrate in three categories: (1)~hedged responses with mixed signals (e.g., ``It was decent but had some issues''), where the embedding is pulled toward multiple anchors simultaneously; (2)~domain-specific language not well-represented in the naturalistic anchors, particularly for ease and importance where the constructs are harder to express as distinctive behavioral experiences; and (3)~responses using understatement or indirectness, where the literal semantic content diverges from the intended sentiment intensity.

\subsubsection{Leave-One-Domain-Out Cross-Validation}

To test whether temperature calibration transfers across semantic domains, we performed leave-one-domain-out (LODO) cross-validation. For each of 8 folds, we calibrated $\tau$ on 7 domains and evaluated on the held-out domain (Table~\ref{tab:cv-lodo}).

\begin{table}[htbp]
\centering
\caption{Leave-one-domain-out cross-validation. $\tau^{*}$: temperature calibrated on training domains only. Aggregate: test predictions pooled across all 8 folds.}
\label{tab:cv-lodo}
\begin{tabular}{lccccc}
\toprule
Held-out domain & $\tau^{*}$ & Train exact & Test exact & Test $\pm$1 & Test MAE \\
\midrule
Satisfaction & 0.05 & 68\% & 5/10 (50\%) & 10/10 (100\%) & 0.50 \\
Likelihood & 0.15 & 66\% & 6/8 (75\%) & 8/8 (100\%) & 0.25 \\
Agreement & 0.15 & 64\% & 7/8 (88\%) & 8/8 (100\%) & 0.13 \\
Ease & 0.15 & 68\% & 5/9 (56\%) & 6/9 (67\%) & 0.78 \\
Importance & 0.15 & 72\% & 3/9 (33\%) & 6/9 (67\%) & 1.00 \\
Trust & 0.15 & 68\% & 5/9 (56\%) & 9/9 (100\%) & 0.44 \\
Value & 0.15 & 67\% & 5/8 (63\%) & 8/8 (100\%) & 0.38 \\
Purchase intent & 0.15 & 64\% & 7/8 (88\%) & 8/8 (100\%) & 0.13 \\
\midrule
\textbf{Aggregate} & & & \textbf{43/69 (62\%)} & \textbf{63/69 (91\%)} & \textbf{0.46} \\
\bottomrule
\end{tabular}
\end{table}

The calibrated temperature is remarkably stable across folds ($\text{mean} = 0.14$, $\text{SD} = 0.03$), with 7 of 8 folds selecting $\tau = 0.15$. This confirms that temperature generalizes across semantic domains: it is a property of the embedding space, not of the specific constructs being measured.

The LODO aggregate (62\% exact, 91\% $\pm$1) is slightly lower than the global evaluation at $\tau = 0.15$ (67\% exact), reflecting the mild disadvantage of using a potentially suboptimal temperature for each test fold. Importantly, the within $\pm$1 rate is nearly identical (91\% vs.\ 91\%), indicating that temperature miscalibration affects precision but not the framework's ability to identify the correct scale region.

\subsubsection{Bootstrap Confidence Intervals}

To estimate confidence intervals for expected performance on new domains, we performed random 70/30 train/test bootstrap cross-validation (100 repeats, seed = 42; Table~\ref{tab:cv-bootstrap}).

\begin{table}[htbp]
\centering
\caption{Bootstrap 70/30 cross-validation (100 repeats). Temperature recalibrated on each training split.}
\label{tab:cv-bootstrap}
\begin{tabular}{lcc}
\toprule
Metric & Mean $\pm$ SD & 95\% CI \\
\midrule
Exact match (\%) & $65.2 \pm 9.0$ & [48, 81] \\
Within $\pm$1 (\%) & $90.9 \pm 5.1$ & [81, 100] \\
MAE & $0.44 \pm 0.12$ & [0.19, 0.67] \\
Calibrated $\tau$ & $0.14 \pm 0.04$ & [0.05, 0.20] \\
\bottomrule
\end{tabular}
\end{table}

\begin{figure}[htbp]
\centering
\includegraphics[width=\textwidth]{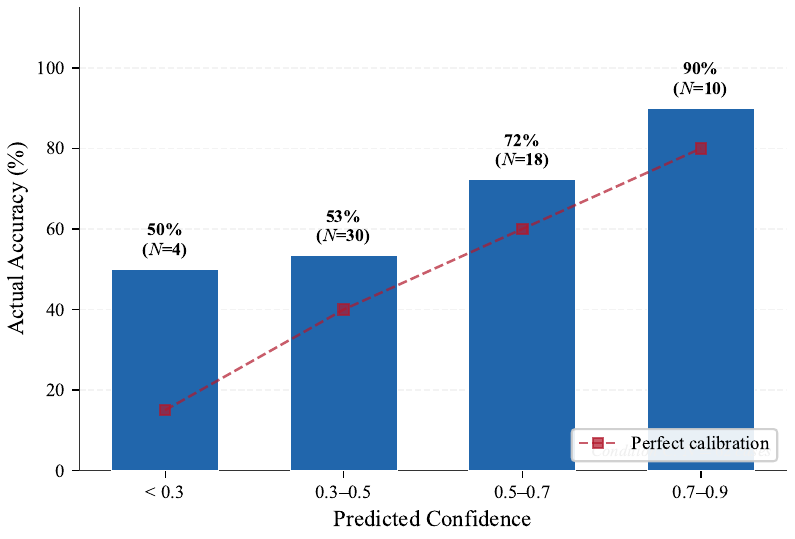}
\caption{Confidence calibration. Predicted confidence (normalized Shannon entropy) vs.\ actual accuracy within confidence bins. Higher confidence predictions are more frequently correct, indicating that the confidence measure provides useful signal about mapping certainty.}
\label{fig:confidence}
\end{figure}

The 95\% confidence interval for exact match (48--81\%) is wide, reflecting both domain-specific performance variation and the moderate size of the test set. However, the within $\pm$1 interval (81--100\%) is more reassuring: even in unfavorable splits, the framework achieves at least 81\% within $\pm$1 accuracy. Figure~\ref{fig:confidence} shows that the confidence measure (normalized Shannon entropy) is well-calibrated: higher predicted confidence corresponds to higher actual accuracy.

\subsubsection{Expanded Ablation Confirmation}

To verify that the factorial findings generalize, we replicated the $2 \times 2$ ablation study (Experiment~3) on the expanded 69-case test set (Table~\ref{tab:ablation-v2}).

\begin{table}[htbp]
\centering
\caption{$2 \times 2$ ablation replicated on expanded test set ($N = 69$, 8 domains). Findings from the initial 17-case ablation are confirmed.}
\label{tab:ablation-v2}
\begin{tabular}{llcccc}
\toprule
Condition & Context & Asymmetric & Exact & Within $\pm$1 & MAE \\
\midrule
A: baseline & No & No & 42/69 (61\%) & 66/69 (96\%) & 0.43 \\
B: H2 only & Yes & No & 37/69 (54\%) & 64/69 (93\%) & 0.55 \\
\textbf{C: H3 only} & \textbf{No} & \textbf{Yes} & \textbf{45/69 (65\%)} & 63/69 (91\%) & \textbf{0.43} \\
D: H2+H3 & Yes & Yes & 43/69 (62\%) & 64/69 (93\%) & 0.45 \\
\bottomrule
\end{tabular}
\end{table}

The expanded ablation confirms all three findings from the initial study: (1) asymmetric embedding (C) produces the highest exact match (65\% vs.\ 61\% baseline, $+4$~pp); (2) contextualization (B) harms accuracy (54\%, $-7$~pp from baseline); and (3) the interaction (D) partially cancels the asymmetric benefit. The effect sizes are attenuated on the harder expanded test set but remain directionally consistent, providing evidence that the factorial findings are not artifacts of the initial 3-domain calibration set.

An unexpected finding is that the baseline achieves the highest within $\pm$1 rate (96\%) despite lower exact match. This suggests that symmetric embedding produces more conservative predictions clustered near the center of the scale, which are more often ``close enough'' even when not exact. Asymmetric embedding trades this conservatism for sharper, more decisive predictions that are more often exactly right but occasionally further off.

\subsection{Baseline Comparison: Direct LLM Rating}
\label{sec:baseline}

To empirically assess whether decoupling generation from measurement improves accuracy, we compared SSR against the standard circular approach: having the same LLM that generated the text also rate it on the Likert scale. We tested two LLM baseline variants to ensure a fair comparison:

\begin{itemize}
\item \textbf{As-deployed}: Uses the exact fallback prompt from our production system (\texttt{mapTo\-Likert\-Via\-LLM}), which provides generic scale labels (``Strongly Disagree/Very Negative'' to ``Strongly Agree/Very Positive'') regardless of the question's semantic domain.
\item \textbf{Domain-aware}: Modifies the prompt to include the question's actual scale anchors (e.g., ``Very dissatisfied'' to ``Very satisfied'' for satisfaction items), giving the LLM domain-specific scale context that SSR also receives through its anchor statements.
\end{itemize}

Both variants use Claude Haiku 4.5 with temperature $= 0$ for deterministic output. Determinism was verified by running the as-deployed variant twice and confirming identical predictions across all 69 cases. This prompt was applied to all 69 reference test cases.\footnote{The baseline experiment script (\texttt{llm-baseline-experiment.ts}) is included in the project repository and can be reproduced with a valid API key.} The results are compared in Table~\ref{tab:baseline}.

\begin{table}[htbp]
\centering
\caption{SSR (embedding-based) vs.\ direct LLM rating on the same 69 test cases. LLM variants use Claude Haiku 4.5 with temperature $= 0$. The as-deployed variant replicates the production fallback; the domain-aware variant provides question-specific scale anchors. Cross-model results (GPT-4o, OpenAI embeddings) are reported in Section~\ref{sec:cross-model}.}
\label{tab:baseline}
\footnotesize
\begin{tabular}{lcccc}
\toprule
Method & Model & Exact & Within $\pm$1 & MAE \\
\midrule
\textbf{LLM baseline (as-deployed)} & \textbf{Claude Haiku 4.5} & \textbf{60/69 (87\%)} & \textbf{69/69 (100\%)} & \textbf{0.13} \\
LLM baseline (domain-aware) & Claude Haiku 4.5 & 60/69 (87\%) & 69/69 (100\%) & 0.13 \\
SSR (embedding) & Voyage 3.5-lite & 45/69 (65\%) & 63/69 (91\%) & 0.43 \\
\bottomrule
\end{tabular}
\normalsize
\end{table}

Both LLM variants achieve identical global accuracy (87\% exact, 100\% within $\pm$1, MAE $= 0.13$), though they differ in which cases they miss: the as-deployed variant errs on 9 cases while the domain-aware variant errs on a different set of 9 cases. All LLM errors across both variants are off-by-one, and 8 of 9 errors in each variant are underestimates (predicted one point below expected), indicating a slight negative bias for ambiguous cases. The convergence in global performance despite different error patterns suggests that scale label specificity does not substantially affect LLM rating accuracy; the model's sentiment analysis capability is robust to label framing.

\subsubsection{Information Asymmetry: Control Condition}

An important methodological difference between the two approaches is that the LLM baseline receives the full question text as part of its prompt, while SSR does not incorporate question text during similarity computation. To quantify whether this information asymmetry explains the accuracy gap, we ran a control condition: both models (Claude Haiku 4.5 and GPT-4o) rating the same 69 test cases with the question text \textit{removed} from the prompt, providing only the response text and scale anchors (the same information available to SSR). Table~\ref{tab:no-question} reports the results.

\begin{table}[htbp]
\centering
\caption{Information asymmetry control: LLM rating accuracy with and without question text on the 69-case benchmark. Removing the question text does not reduce accuracy; in fact, both models improve slightly.}
\label{tab:no-question}
\begin{tabular}{llccc}
\toprule
Model & Condition & Exact & Within $\pm$1 & MAE \\
\midrule
Claude Haiku 4.5 & with question & 60/69 (87\%) & 69/69 (100\%) & 0.13 \\
Claude Haiku 4.5 & \textbf{no question} & \textbf{62/69 (90\%)} & 69/69 (100\%) & 0.10 \\
GPT-4o & with question & 55/69 (80\%) & 69/69 (100\%) & 0.20 \\
GPT-4o & \textbf{no question} & \textbf{59/69 (86\%)} & 69/69 (100\%) & 0.14 \\
\midrule
SSR (Voyage) & no question* & 45/69 (65\%) & 63/69 (91\%) & 0.43 \\
\bottomrule
\multicolumn{5}{l}{\small * SSR never receives question text by design.} \\
\end{tabular}
\end{table}

The result is the opposite of what an information-asymmetry explanation would predict: removing the question text \textit{improves} accuracy for both models (Claude: $+3$~pp; GPT-4o: $+6$~pp). This is consistent with the ablation finding that question context can introduce shared semantic content that degrades scale discrimination (Section~\ref{sec:ablation}, H2 contextualization). The question text appears to slightly bias the LLM toward the question's framing rather than the response's content, producing miscalibrated ratings on ambiguous cases.

This finding has an important implication: \textbf{SSR's lower accuracy (65\%) relative to the LLM baseline (87--90\%) cannot be attributed to missing information}. Even under equal-information conditions (no question text), the LLM outperforms SSR by 21--25~pp. The accuracy gap therefore reflects a fundamental difference in rating capability: autoregressive language models have internalized ordinal scale mapping from their training data, while embedding models must reconstruct this mapping from geometric similarity alone.

These results have three important implications. First, they demonstrate that modern LLMs are highly skilled at sentiment-to-scale mapping even without question context, consistent with deeply internalized ordinal scale representations from pretraining. Second, they confirm that our reference ratings are \textit{not} idiosyncratic: a general-purpose LLM converges on the same scale point in 87--90\% of cases, suggesting that the test cases capture recognizable patterns of Likert-level sentiment. Third, and most critically for SSR's positioning, they establish that \textbf{SSR's contribution is methodological independence, not accuracy superiority}. The value of decoupling generation from measurement lies not in producing more accurate ratings, but in providing ratings from a fundamentally different computational mechanism, one that cannot share the generative model's biases, training artifacts, or internal consistency patterns.

A further limitation of this comparison merits emphasis: our benchmark tests the LLM on \textit{human-written} texts, not on its own output. This is necessary for controlled evaluation (both methods rate the same text), but it does not capture the full circularity scenario, where the same model generates a text and then evaluates it. Self-evaluation may introduce additional biases (e.g., inflated confidence in the coherence of its own output) that would not manifest when rating human-written text. Experiment~4 (Section~\ref{sec:circularity}) directly tests this scenario.

\subsection{Cross-Model Generalization}
\label{sec:cross-model}

The preceding experiments used a single embedding model (Voyage 3.5-lite) and a single LLM family (Anthropic Claude) for baseline comparison. To assess whether our findings generalize across model providers, we evaluated both the SSR framework and the LLM baseline using alternative models: OpenAI \texttt{text-embedding-3-small} (1536 dimensions) for embedding-based mapping and GPT-4o (temperature $= 0$) for direct LLM rating. This directly addresses the concern that SSR's accuracy or the baseline comparison could be artifacts of a specific model's characteristics, and follows the multi-model approach of \citet{maier2024ssr}, who validated SSR with GPT-4o, Gemini-2f, and OpenAI embeddings.

\subsubsection{Embedding Model Comparison}

Because OpenAI's embedding API does not support asymmetric embedding (there is no \texttt{input\_type} parameter), we compare Voyage symmetric against OpenAI symmetric to ensure an apples-to-apples evaluation. Voyage asymmetric results are included as a Voyage-specific optimization. Table~\ref{tab:cross-embedding} reports results at the best temperature for each provider.\looseness=-1

\begin{table}[htbp]
\centering
\caption{SSR accuracy across embedding models on the 69-case benchmark. Voyage symmetric vs.\ OpenAI symmetric is the fair cross-model comparison; Voyage asymmetric is a provider-specific optimization. All use min-max normalization.}
\label{tab:cross-embedding}
\begin{tabular}{llcccc}
\toprule
Embedding Model & Mode & Best $\tau$ & Exact & Within $\pm$1 & MAE \\
\midrule
Voyage 3.5-lite & asymmetric & 0.15 & 46/69 (67\%) & 63/69 (91\%) & 0.42 \\
Voyage 3.5-lite & symmetric & 0.20 & 45/69 (65\%) & 66/69 (96\%) & 0.41 \\
\textbf{OpenAI emb-3-small} & \textbf{symmetric} & \textbf{0.20} & \textbf{53/69 (77\%)} & \textbf{69/69 (100\%)} & \textbf{0.23} \\
\bottomrule
\end{tabular}
\end{table}

\begin{figure}[htbp]
\centering
\includegraphics[width=0.95\textwidth]{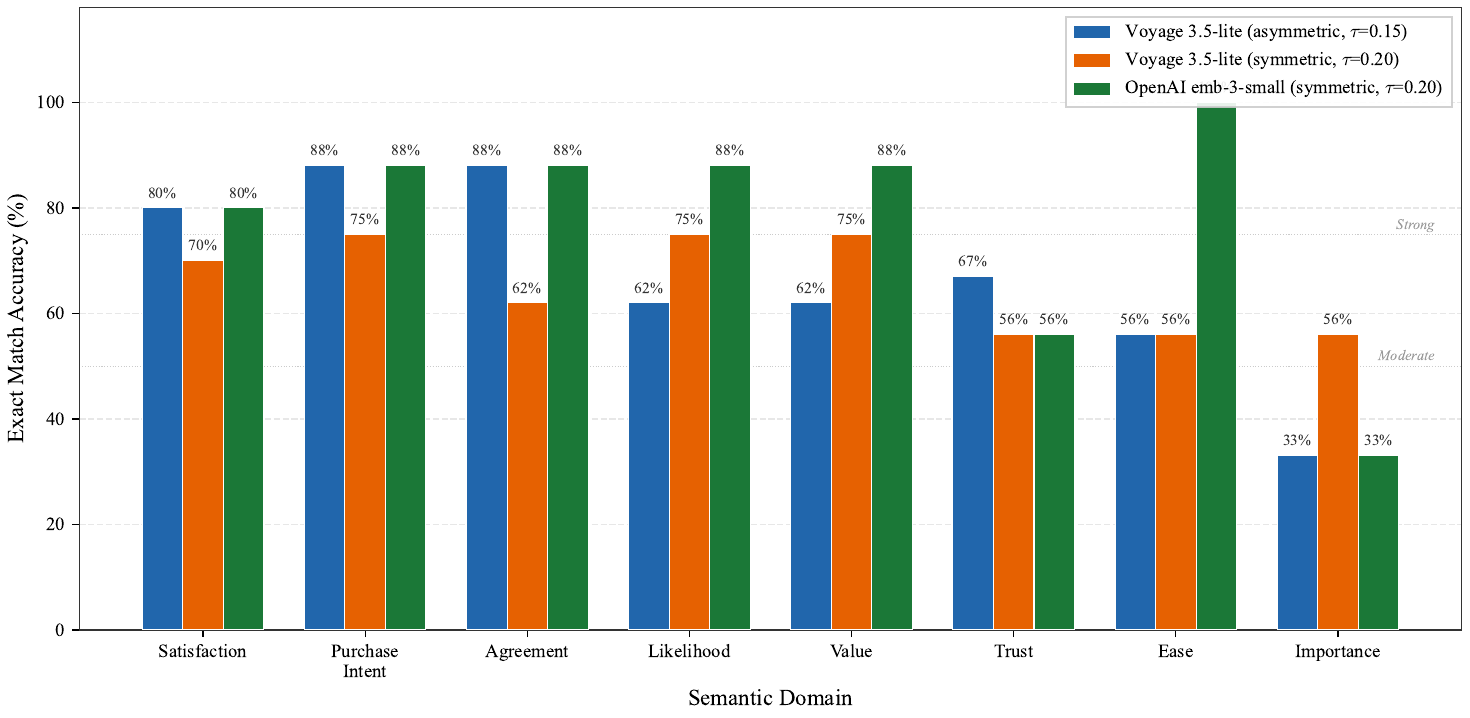}
\caption{Per-domain exact match accuracy across embedding models. OpenAI \texttt{text-embedding-3-small} (green) outperforms both Voyage configurations on most domains. The largest improvement is on ease (56\% $\to$ 100\%), while importance remains weak across all models (33\%), confirming that this is an anchor quality issue rather than an embedding model limitation. Domains sorted by descending Voyage asymmetric accuracy.}
\label{fig:cross-model}
\end{figure}

OpenAI's embedding model substantially outperforms Voyage in symmetric mode (+12~pp exact match, +4~pp within $\pm$1, $-0.18$ MAE). Figure~\ref{fig:cross-model} visualizes the per-domain comparison: OpenAI achieves 100\% exact match on ease (the weakest Voyage domain) and 88\% on likelihood, agreement, value, and purchase intent.

The improvement is most pronounced on domains where Voyage struggled (ease: 56\% $\to$ 100\%, likelihood: 63\% $\to$ 88\%), suggesting that OpenAI's 1536-dimensional space provides better discrimination for constructs that Voyage's 1024-dimensional space compresses.

The persistence of the importance domain as the weakest (33\% for both providers) indicates that this is an anchor quality issue, not an embedding model issue: the importance anchor statements do not yet provide sufficient semantic discrimination regardless of the embedding space.

Critically, OpenAI achieves 100\% within $\pm$1 across all five temperatures tested ($\tau$ from 0.05 to 0.25), compared to Voyage symmetric's 88--97\%. This indicates that the SSR framework's ability to identify the correct scale \textit{region} generalizes robustly across embedding providers.

\subsubsection{LLM Baseline Comparison}

We replicated the baseline comparison (Section~\ref{sec:baseline}) using GPT-4o alongside Claude Haiku 4.5, both at temperature $= 0$. Table~\ref{tab:cross-baseline} reports results for all four model--variant combinations.

\begin{table}[htbp]
\centering
\caption{LLM baseline accuracy across model families. Both models tested with generic labels (as-deployed) and question-specific labels (domain-aware) on the same 69 test cases.}
\label{tab:cross-baseline}
\begin{tabular}{llccc}
\toprule
Model & Variant & Exact & Within $\pm$1 & MAE \\
\midrule
Claude Haiku 4.5 & as-deployed & 60/69 (87\%) & 69/69 (100\%) & 0.13 \\
Claude Haiku 4.5 & domain-aware & 60/69 (87\%) & 69/69 (100\%) & 0.13 \\
GPT-4o & as-deployed & 57/69 (83\%) & 69/69 (100\%) & 0.17 \\
GPT-4o & domain-aware & 55/69 (80\%) & 69/69 (100\%) & 0.20 \\
\bottomrule
\end{tabular}
\end{table}

Both models achieve high accuracy ($\geq 80$\% exact, 100\% within $\pm$1), confirming that sentiment-to-scale mapping is a well-learned capability across model families. Cross-model agreement is 81\% (56/69 same prediction on the as-deployed variant), with each model outperforming the other on different cases: Claude is uniquely correct on 8 cases, GPT-4o on 5, and both miss 4. All errors across both models are off-by-one, consistent with boundary ambiguity in the reference ratings rather than systematic model failure.

\paragraph{Label sensitivity.}
A notable divergence emerges in label sensitivity. Claude exhibits complete \textit{label invariance}: both variants produce identical global accuracy (87\%), though they differ on which specific cases they miss. GPT-4o, in contrast, shows a 3-percentage-point accuracy decrease with domain-aware labels (83\% $\to$ 80\%), suggesting that question-specific scale anchors introduce slight miscalibration in GPT-4o's scale mapping. This is the opposite of what one might expect (more specific labels should help) and suggests that GPT-4o's internal scale representation is well-calibrated for generic evaluative language but slightly disrupted by domain-specific terminology. This finding nuances the label invariance observation from Section~\ref{sec:baseline}: label invariance holds for Claude but may not generalize to all model families.

\paragraph{Implications.}
The cross-model results strengthen both the robustness of the accuracy--independence tradeoff and the potential for SSR accuracy improvement through embedding model selection without compromising methodological independence.

\subsection{Experiment 4: Self-Rating Circularity Test}
\label{sec:circularity}

\subsubsection{Motivation}

The baseline comparison (Section~\ref{sec:baseline}) established that direct LLM rating achieves higher accuracy than SSR on human-written text. However, that comparison does not test the circularity scenario that motivates SSR: a model rating \textit{its own} generated text. When the same model generates and evaluates a response, the evaluation may reflect the model's internal consistency, its tendency to produce text that aligns with its own classification boundaries, rather than the text's true semantic content. This experiment directly tests whether self-rating produces measurable bias.

\subsubsection{Design}

We generated synthetic survey responses for all 69 reference test cases using Claude Haiku 4.5 with LLM generation temperature $T = 0.7$ (distinct from the SSR softmax temperature $\tau$) and five persona profiles varying in communication style (detailed/analytical, blunt/direct, hedging/qualified, practical/concise, and informal/expressive). Each persona generated responses at all target rating levels (1--5), producing $69 \times 5 = 345$ generated texts. Each text was then independently rated by two methods:

\begin{itemize}
  \item \textbf{LLM self-rating} (circular): Claude Haiku 4.5 with $T = 0$, using the as-deployed prompt from Section~\ref{sec:baseline}.
  \item \textbf{SSR} (independent): The final configuration (asymmetric embedding, naturalistic anchors, min-max normalization, $\tau = 0.15$).
\end{itemize}

The generation prompt provided sentiment guidance derived from the question's scale anchors (e.g., ``Very dissatisfied'' to ``Very satisfied''), not SSR's naturalistic anchor statements, to prevent lexical contamination that could artificially inflate SSR accuracy. All hypotheses, metrics, and statistical tests were pre-registered before data collection.\footnote{Pre-registered via timestamped git commit (February 9, 2026). The full pre-registration protocol, including the three pre-written narrative scenarios and all analysis code, was finalized and committed before Phase~1 (text generation) began. The cross-model control (Section~\ref{sec:crossmodel-circularity}) was added post-hoc but all analyses were specified before examining cross-model results.}

This experimental design yields a complete $2 \times 2$ comparison when combined with prior results (Table~\ref{tab:2x2}).

\begin{table}[htbp]
\centering
\caption{$2 \times 2$ design: exact match accuracy (\%) by text source (human vs.\ generated) and rating method (LLM vs.\ SSR). Cond.\ A: LLM rates human text (Section~\ref{sec:baseline}); Cond.\ B: SSR rates human text (Section~\ref{sec:cross-validation}); Cond.\ C: LLM rates its own generated text; Cond.\ D: SSR rates generated text.}
\label{tab:2x2}
\begin{tabular}{lccc}
\toprule
 & Human text & Generated text & $\Delta$ \\
\midrule
LLM rating & 87\% (Cond.\ A, $N=69$) & 73\% (Cond.\ C, $N=345$) & $-14$~pp \\
SSR rating & 65\% (Cond.\ B, $N=69$) & 50\% (Cond.\ D, $N=345$) & $-15$~pp \\
\midrule
$\Delta$ & $+22$~pp & $+23$~pp & \\
\bottomrule
\end{tabular}
\end{table}

Both methods show comparable accuracy degradation when moving from human to generated text ($-14$~pp for LLM, $-15$~pp for SSR). The near-identical magnitude of this degradation ($\Delta\Delta = 1$~pp) suggests that generated text is inherently harder to rate regardless of method; the difficulty is a property of the text, not the rating approach. The LLM's accuracy advantage over SSR is stable across text sources (+22~pp on human text, +23~pp on generated text), indicating no interaction between text source and rating method at the level of aggregate accuracy. However, the \textit{nature} of the errors differs fundamentally, as the confirmatory tests below demonstrate.

\subsubsection{Results}

Five pre-registered confirmatory tests were conducted with Holm--Bonferroni correction for multiple comparisons ($\alpha = 0.05$; Table~\ref{tab:confirmatory}).

\begin{table}[htbp]
\centering
\caption{Pre-registered confirmatory tests ($N = 345$ generated texts). All $p$-values are Holm--Bonferroni corrected. Tests marked *** are significant at the corrected $\alpha = 0.05$ level.}
\label{tab:confirmatory}
\scriptsize
\begin{tabular}{lcccl}
\toprule
Test & Statistic & $p_{\text{adj}}$ & Effect size & Interpretation \\
\midrule
1. Rating divergence & $W(345) = 5814$ & $< .001$*** & $r = .40$ & Significant LLM--SSR divergence \\
2. LLM auto-accuracy & $\chi^2(1) = 4.0$ & $.237$ & OR $= 0.36$ & No sig.\ accuracy change (LLM) \\
3. SSR gen-accuracy & $\chi^2(1) = 1.4$ & $.237$ & OR $= 0.56$ & No sig.\ accuracy change (SSR) \\
4. Directional bias & $W(345) = 5814$ & $< .001$*** & $r = .40$ & Systematic LLM negative bias \\
5. Variance compression & $F(1, 688) = 60.0$ & $< .001$*** & $\sigma^2_\text{LLM}/\sigma^2_\text{SSR} = 0.24$ & 4$\times$ LLM variance compression \\
\bottomrule
\end{tabular}
\normalsize
\end{table}

\begin{figure}[htbp]
\centering
\includegraphics[width=0.95\textwidth]{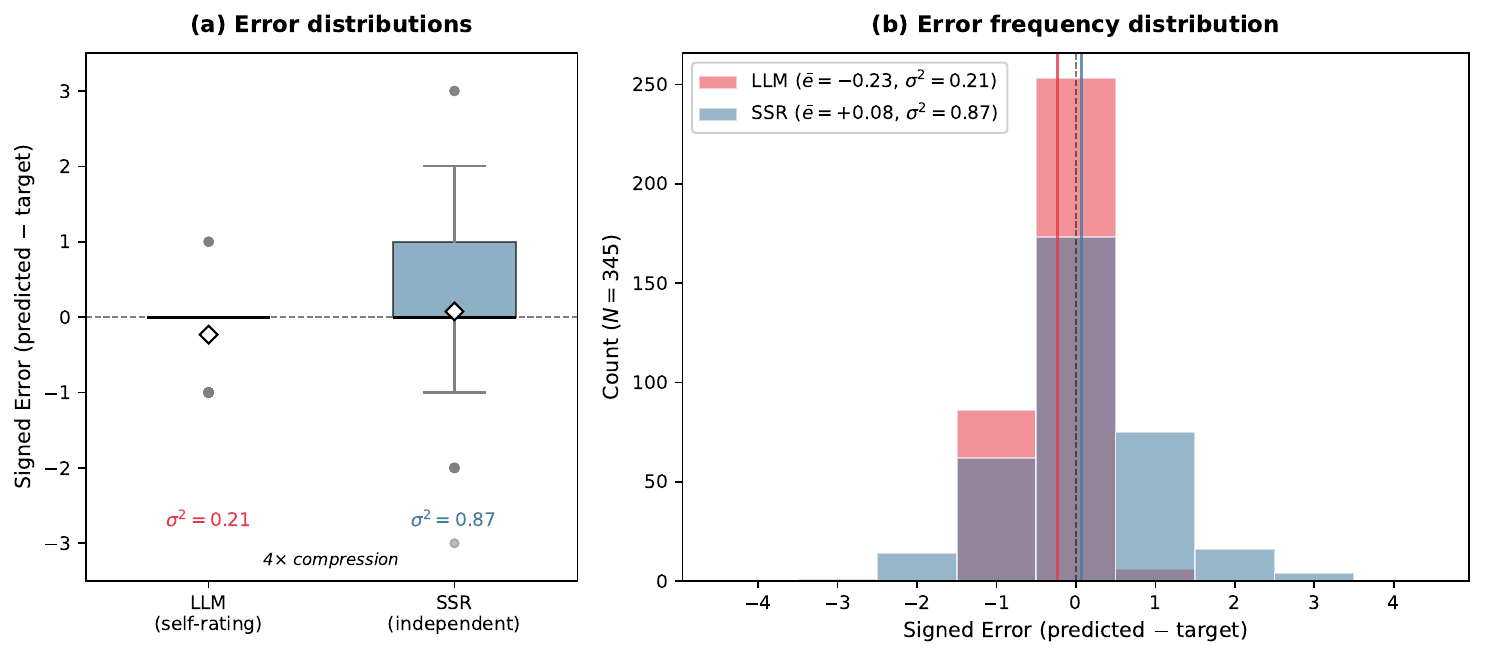}
\caption{Error distribution comparison between LLM self-rating (circular) and SSR (independent) on 345 generated texts. (a)~Box plots show 4-fold variance compression in LLM errors ($\sigma^2 = 0.21$) compared to SSR ($\sigma^2 = 0.87$); diamonds indicate means. (b)~Histograms reveal that LLM errors concentrate at 0 and $-1$ with a systematic negative bias ($\bar{e} = -0.23$), while SSR errors are more broadly distributed but centered near zero ($\bar{e} = +0.08$). Vertical lines indicate mean error for each method. A cross-model control (GPT-4o rating the same texts) confirms that the compression pattern generalizes across model families (Table~\ref{tab:crossmodel}).}
\label{fig:circularity}
\end{figure}

Three of five tests are significant after correction: rating divergence (Test~1), directional bias (Test~4), and variance compression (Test~5). The two accuracy-comparison tests (Tests~2--3) are not significant, indicating that neither method's accuracy changes significantly between human and generated text at the aggregate level. Figure~\ref{fig:circularity} visualizes the key finding: LLM self-rating errors cluster tightly around a small negative offset, while SSR errors are more dispersed but centered near zero.

\paragraph{Rating divergence.}
The mean difference between LLM self-rating and SSR rating across 345 generated texts was $-0.31$ ($\text{SD} = 0.99$; Wilcoxon $W(345) = 5814$, $p < .001$, $r = .40$). The negative sign indicates that LLM ratings are systematically lower than SSR ratings. This medium-to-large effect ($r = .40$) reflects a genuine divergence in how the two methods map the same text to scale points.

\paragraph{Directional bias.}
LLM self-rating shows a consistent downward bias relative to reference ratings (mean signed error $= -0.23$, $\text{SD} = 0.46$), while SSR shows near-zero bias (mean signed error $= +0.08$, $\text{SD} = 0.93$). This pattern, the LLM systematically underestimating the intended rating when evaluating its own generated text, is the opposite of the na\"ive prediction that self-rating would produce \textit{inflated} agreement. We hypothesize that this reflects a temperature asymmetry between generation and evaluation: the generator ($T = 0.7$) samples from a broader distribution that includes more expressive, emotionally vivid language, while the rater ($T = 0$) deterministically selects the most conservative interpretation of that text. In effect, the generator ``writes like a 4'' but the rater ``reads like a 3.'' This is a testable hypothesis: if correct, increasing the rater's temperature should reduce the negative bias (at the cost of increased variance), and decreasing the generator's temperature should produce text that the rater scores more accurately. Testing this hypothesis is a priority for future work but is beyond the scope of the present study.

\paragraph{Variance compression.}
LLM rating errors show 4-fold variance compression: $\sigma^2 = 0.21$ compared to SSR's $\sigma^2 = 0.87$ (Levene $F(1, 688) = 60.0$, $p < .001$). This means LLM errors cluster tightly around a small negative offset, while SSR errors are more dispersed but centered near zero. Crucially, the cross-model control (Section~\ref{sec:crossmodel-circularity}) shows that GPT-4o produces nearly identical compression when rating Claude-generated text ($\sigma^2 = 0.23$, within/cross ratio $= 0.93$), indicating that this effect is a general property of LLM-based rating rather than a consequence of a model evaluating its own output. The compression likely reflects how autoregressive models process evaluative text, converging on similar scale interpretations regardless of the text's origin, rather than shared internal representations between generator and rater.

\paragraph{Self-agreement.}
LLM accuracy on self-generated text (77\% exact when using the modal rating across 5 personas per case, compared to 73\% per-observation in Table~\ref{tab:2x2}) was not significantly different from its accuracy on human-written text (87\%; McNemar $\chi^2 = 4.0$, $p_{\text{adj}} = .237$). While the point estimate drops 10 percentage points, the relatively small number of discordant pairs ($b = 4$, $c = 11$) limits statistical power. SSR accuracy similarly did not change significantly (55\% vs.\ 65\%; McNemar $\chi^2 = 1.4$, $p_{\text{adj}} = .237$).

\subsubsection{Exploratory Analyses}

Table~\ref{tab:circularity-domain} shows per-domain accuracy and divergence on generated text. Two domains exhibit a reversal where SSR outperforms LLM self-rating: likelihood (57\% vs.\ 53\%) and purchase intent (70\% vs.\ 57\%). Notably, purchase intent is the construct for which \citet{maier2024ssr} demonstrated distributional validity against real human data, and it is also a domain where SSR achieves strong performance on human text (88\%, Section~\ref{sec:cross-validation}).

\begin{table}[htbp]
\centering
\caption{Per-domain accuracy and divergence on generated text. Domains sorted by LLM exact match.}
\label{tab:circularity-domain}
\begin{tabular}{lcccc}
\toprule
Domain & LLM exact & SSR exact & Mean divergence & $N$ \\
\midrule
Importance & 84\% & 38\% & $+0.56$ & 45 \\
Agreement & 83\% & 53\% & $-0.38$ & 40 \\
Ease & 82\% & 49\% & $-1.02$ & 45 \\
Value & 80\% & 50\% & $+0.03$ & 40 \\
Trust & 76\% & 53\% & $-0.44$ & 45 \\
Satisfaction & 70\% & 36\% & $-0.56$ & 50 \\
Purchase intent & 57\% & 70\% & $-0.35$ & 40 \\
Likelihood & 53\% & 57\% & $-0.23$ & 40 \\
\bottomrule
\end{tabular}
\end{table}

Additional exploratory findings: (1)~No significant persona$\times$method interaction (Kruskal--Wallis $H = 6.09$, $p = .19$); text complexity drives divergence, with the informal persona (Ryan) showing the smallest $|\bar{d}| = 0.20$ and the detailed persona (Maria) the largest $|\bar{d}| = 0.46$. (2)~LLM errors show near-zero bias at extremes but substantial underestimation at target $= 2$ ($\bar{e} = -0.66$); SSR shows the expected regression-to-the-mean pattern. (3)~SSR confidence was higher on generated text (mean $= 0.57$) than human text (mean $= 0.50$; $U = 14964$, $p < .001$), consistent with generated text being semantically ``cleaner.''

\subsubsection{Cross-Model Circularity Control}
\label{sec:crossmodel-circularity}

To disentangle within-model circularity from general LLM rating properties, we rated all 345 Claude-generated texts with GPT-4o (temperature $= 0$). Table~\ref{tab:crossmodel} reports the three-condition comparison.

\begin{table}[htbp]
\centering
\caption{Three-condition comparison on 345 Claude-generated texts. Within-model and cross-model LLM rating show nearly identical variance compression relative to SSR (within/cross ratio $= 0.93$), indicating that compression is a general LLM property.}
\label{tab:crossmodel}
\begin{tabular}{lcccccc}
\toprule
Condition & Exact & $\pm$1 & MAE & $\bar{e}$ & SD & $\sigma^2$ \\
\midrule
Within-model (Claude$\to$Claude) & 73\% & 100\% & 0.27 & $-0.23$ & 0.46 & 0.213 \\
Cross-model (Claude$\to$GPT-4o) & 75\% & 100\% & 0.25 & $-0.15$ & 0.48 & 0.229 \\
Independent (Claude$\to$SSR) & 50\% & 90\% & 0.61 & $+0.08$ & 0.93 & 0.870 \\
\bottomrule
\end{tabular}
\end{table}

The cross-model condition shows nearly identical variance to within-model ($\sigma^2 = 0.229$ vs.\ $0.213$; ratio $= 0.93$), with 86\% agreement between Claude and GPT-4o (296/345 identical predictions). Both LLM conditions show approximately 4-fold compression relative to SSR. The cross-model condition shows slightly reduced directional bias ($\bar{e} = -0.15$ vs.\ $-0.23$), suggesting a small within-model component, though the bias remains negative and substantially different from SSR's near-zero bias ($+0.08$).

This resolves a key interpretive ambiguity: variance compression is a general property of LLM-based rating, not a within-model circularity artifact. The implication is that \textbf{any LLM-based measurement pipeline will exhibit compressed error distributions relative to embedding-based measurement}, regardless of whether the generation and rating models are the same. SSR's higher variance ($\sigma^2 = 0.87$) reflects genuine measurement uncertainty from a geometrically independent mechanism, with near-zero mean bias making errors amenable to population-level cancellation.


\section{Discussion}
\label{sec:discussion}

\subsection{Key Findings}

Our calibration study yielded several principal findings, each with methodological implications extending beyond the specific application of synthetic survey generation.

\subsubsection{Circularity Can Be Substantially Reduced: At a Cost}

The SSR framework demonstrates that text-to-scale mapping can be performed by a measurement channel that is architecturally independent from the generation model. The embedding model (Voyage AI) operates on a fundamentally different principle than the autoregressive generation model (Claude): it maps text to geometric positions in a continuous vector space and computes similarity via dot products, without any ``understanding'' of scale semantics or survey methodology. We emphasize that \textit{architectural} independence does not guarantee \textit{statistical} independence: because embedding models and LLMs are trained on overlapping web-scale corpora, residual correlations between their representations are plausible. SSR substantially reduces circularity by using a different computational mechanism, but does not eliminate all shared information between the generation and measurement channels.

However, our baseline comparison (Table~\ref{tab:baseline}) reveals that this independence comes at a substantial accuracy cost. This gap is not surprising in retrospect. Autoregressive language models are trained on billions of human-annotated examples that implicitly encode the very mappings SSR attempts to reconstruct from scratch via geometric similarity. SSR, by design, discards this learned knowledge in favor of independence.

The appropriate interpretation depends on the research context. When the goal is \textit{maximizing accuracy}, direct LLM rating is more accurate. When the goal is \textit{methodological independence}, SSR provides a guarantee that no amount of LLM accuracy can substitute. These are complementary, not competing, objectives. The cross-model comparison (Section~\ref{sec:cross-model}) confirms this tradeoff is robust across model families, and the circularity experiment with cross-model control (Section~\ref{sec:circularity}) demonstrates that LLM-based rating produces systematically different error patterns than independent measurement, not just within-model, but across model families.

\subsubsection{Anchor Quality Dominates All Other Factors}

The dominance of anchor quality over all algorithmic manipulations connects directly to the survey methodology literature on behavioral anchoring \citep{krosnick2010question, tourangeau2000psychology}: just as human respondents are more reliable with concrete behavioral scale anchors, embedding models produce more discriminative vectors when anchors contain distinctive, specific language. The mechanism is analogous: richer semantic content provides better ``targets'' for mapping, whether the mapping is cognitive (human) or geometric (embedding). This principle extends beyond SSR to any embedding-based text classification against semantic prototypes.

\subsubsection{Mapping Error Magnitude in Context}

The accuracy gap reflects a fundamental difference in rating mechanism: autoregressive models have internalized ordinal mapping from pretraining, while embedding models must reconstruct it geometrically. The information asymmetry control (Table~\ref{tab:no-question}) rules out the hypothesis that LLMs simply have access to more information. Crucially, the gap is concentrated in weaker domains; for well-anchored domains (satisfaction, purchase intent), all methods achieve $\geq$80\% exact. The 12~pp improvement from Voyage to OpenAI embeddings, without any change to the framework, demonstrates that model selection is a high-leverage path that does not compromise independence.

For context, human test-retest reliability on 5-point Likert scales typically produces response standard deviations of 0.5--1.0 \citep{nunnally1994psychometric, krosnick2010question}, though test-retest measures \textit{intra-rater stability} while our evaluation measures \textit{inter-method agreement} (see Limitations).

\paragraph{Connection to distributional validation.}
Our purchase intent results (88\% exact, MAE $= 0.13$) converge with \citeauthor{maier2024ssr}'s (\citeyear{maier2024ssr}) distributional validation (KS similarity $> 0.85$) for the same construct, providing indirect cross-study evidence that SSR generalizes across research groups, embedding models, and evaluation methodologies. However, individual-level accuracy and distributional validity measure fundamentally different properties: a method with 65\% individual exact match can still produce excellent distributional fit if errors are symmetric, while a method with high individual accuracy but systematic bias would fail distributional validation. Bridging these two levels, case-level precision and population-level fidelity, is a priority for future work.

\subsubsection{Label-Invariant LLM Accuracy}

Label invariance, identical LLM accuracy regardless of scale label formulation, deepens the circularity concern: if LLMs have internalized ordinal mapping as a general capability, then text generation and scale assignment draw on the same learned construct. The cross-model comparison reveals this is not universal (Claude: 87\% invariant; GPT-4o: 3~pp decrease with domain-aware labels), suggesting that the severity of internalized psychometric knowledge varies across model families. A practical implication is that \textbf{the rating step does not require expensive models}: Claude Haiku 4.5 reaches the estimated arguability ceiling ($\sim$87--90\%; Section~\ref{sec:arguability}).

\subsubsection{SSR as a Validation Layer}

The accuracy--independence tradeoff suggests a practical hybrid architecture: use the LLM for primary rating (maximizing accuracy) and SSR as an independent validation check. Cases where the two methods disagree by more than one scale point can be flagged for manual review or assigned lower confidence. The cross-model circularity control (Section~\ref{sec:crossmodel-circularity}) strengthens this recommendation: since variance compression is a general LLM property (not just within-model), \textit{any} LLM-based rating pipeline will underestimate measurement uncertainty. SSR provides a calibration reference with genuine measurement noise, enabling researchers to detect when LLM ratings are artificially precise.

\subsection{Limitations}
\label{sec:limitations}

\subsubsection{Single-Rater Ground Truth}

This is the most significant limitation of the present work. All expected ratings were assigned by a single researcher, not derived from multi-rater coding or crowd-sourced consensus. Typical inter-rater reliability for Likert-scale coding ranges from $\kappa = 0.60$--$0.80$ \citep{landis1977measurement}, and for subtle cases (e.g., mixed-signal responses near the boundary between ratings 2 and 3), reasonable coders might disagree on the ``correct'' rating. Our reported accuracy metrics therefore conflate SSR error with the uncertainty in the ground truth itself. A multi-rater validation study with established inter-rater agreement (e.g., Krippendorff's $\alpha \geq 0.80$) is planned for the journal version of this paper and would provide more robust benchmarks for SSR's true accuracy.

\subsubsection{Calibration Set Size and Domain Coverage}

Our initial optimization was conducted on 17 test cases across 3 domains. The expanded 69-case set with leave-one-domain-out cross-validation provides more robust generalization evidence, but also reveals substantial domain-specific variation (33--90\% exact match). The weak performance on ease and importance suggests that the naturalistic anchor statements for these constructs may not yet provide sufficient embedding discrimination. A full validation study against real human data remains essential for definitive generalization claims.

\subsubsection{Domain-Specific Anchor Quality}

The cross-validation reveals that not all naturalistic anchor families are equally effective. Satisfaction and purchase intent anchors achieve 88--90\% exact match, while ease and importance achieve only 33--56\%. This may reflect the inherent difficulty of expressing ease and importance as distinctive behavioral experiences, or it may indicate that these specific anchor statements require further iterative refinement. The framework's modularity, since anchor families can be updated independently, supports iterative improvement.

\subsubsection{English and Consumer Domain Only}

All test cases are in English and concern consumer experiences (e-commerce, product quality, purchase intent). The anchor quality finding may not transfer to other languages, where embedding models may have different similarity compression properties. Similarly, domains with more complex response patterns (e.g., political attitudes, health surveys, organizational behavior) may present challenges not captured in our test set.

\subsubsection{Two Embedding Models Tested}

We validated SSR with two commercial embedding models (Voyage AI \texttt{voyage-3.5-lite} and OpenAI \texttt{text-embedding-3-small}), finding that both produce viable results with OpenAI achieving substantially higher accuracy (77\% vs.\ 67\% exact). While this demonstrates cross-provider generalization, both are proprietary cloud-hosted models. Testing with open-source alternatives (E5, BGE, GTE) and locally-hosted models would further strengthen generalizability claims and enable deployment without API dependencies. Additionally, higher-dimensional models such as OpenAI \texttt{text-embedding-3-large} (3072 dimensions) may offer even better discrimination.

\subsubsection{Circularity Experiment Scope}

The circularity experiment uses texts generated by a single LLM (Claude Haiku 4.5) and one cross-model control (GPT-4o). While the near-identical compression across these two families is suggestive, confirming generality requires testing with additional generators and raters. The cross-model control was added after the initial pre-registration; we note this addition transparently, though all analyses were specified before examining cross-model results.

\subsubsection{Ground Truth Arguability}
\label{sec:arguability}

Approximately 14 of the 69 test cases ($\sim$20\%) have expected ratings that a reasonable second rater might assign differently, typically to an adjacent scale point. These 14 cases were identified by the first author during test set construction, \textit{before} any model experiments were conducted, by flagging cases where the difficulty label was ``subtle'' or ``edge'' and the response text contained mixed signals that could plausibly support an adjacent rating. This assessment is a self-evaluation of one's own uncertainty rather than an independent inter-rater judgment, and the true arguability rate may differ from this estimate. Nonetheless, the concentration in ``subtle'' and ``edge'' categories places a theoretical ceiling on exact-match accuracy of approximately 87--90\%: any method achieving this level is effectively at the limit of single-rater ground truth reliability. The LLM baseline's 87\% exact match approaches this ceiling, consistent with both the LLM's strong sentiment analysis and the inherent ambiguity in the ground truth itself.

Importantly, we did not revise expected ratings post-hoc based on model performance. Doing so would constitute data dredging and would invalidate the evaluation. The reference ratings were fixed before any experiments were conducted. Multi-rater validation, using independent coders with established inter-rater agreement, is planned for the journal version and would allow distinguishing genuine model errors from reference rating ambiguity.

\subsection{Implications for Practice}

\subsubsection{For Researchers Using Synthetic Survey Data}

\begin{enumerate}
\item \textbf{Consider the accuracy--independence tradeoff}: Direct LLM rating is more accurate but does not provide independent measurement. For exploratory analyses where accuracy is paramount, LLM rating may be appropriate. For confirmatory studies or psychometric validation, SSR or similar decoupled approaches should be preferred. Reporting the measurement method is essential regardless.

\item \textbf{Invest in anchor quality}: When using embedding-based classification, the quality of your prototype descriptions matters more than algorithm tuning. Write concrete, distinctive, emotionally rich descriptions rather than abstract labels.

\item \textbf{Report measurement methodology}: Papers using synthetic survey data should clearly specify how text-to-scale mapping is performed, just as papers using human surveys report their scale design and administration procedures.
\end{enumerate}

\subsubsection{For the Computational Social Science Community}

The SSR framework is released as open-source software with full calibration data (15 anchor families, 69-case benchmark, experiment scripts, multi-model comparison results), enabling replication and extension. Three concrete extensions would be most valuable: (1)~translating the anchor library to other languages and testing cross-lingual embedding performance; (2)~benchmarking SSR with additional embedding models (Cohere, open-source E5/BGE/GTE) and larger-dimensional variants beyond the two providers tested here; and (3)~contributing anchor families for domains not yet covered (e.g., clinical, organizational, political).

\subsection{Future Work}

Several extensions are planned or in progress:

\begin{enumerate}
\item \textbf{Distributional validation against human data}: \citet{maier2024ssr} demonstrated that SSR achieves 90\% of human test--retest reliability at the population level, but their evaluation and ours use different embedding models, anchor strategies, and evaluation metrics. The critical next step is to apply our calibrated SSR configuration (naturalistic anchors, asymmetric embedding) to real survey instruments with human response data, enabling direct distributional comparison (KS tests, distribution shape analysis) alongside the individual-level metrics reported here. This would bridge the case-level precision established in the present work with the population-level fidelity demonstrated by \citeauthor{maier2024ssr}, providing a complete validation picture.
\item \textbf{Anchor refinement for weak domains}: The 22-percentage-point accuracy gap between SSR and LLM baseline is concentrated in weaker domains. Iterative improvement of ease, importance, and trust anchors, guided by the per-domain analysis in Section~\ref{sec:cross-validation}, is the highest-leverage path to closing this gap.
\item \textbf{Broader multi-model comparison}: Our cross-model comparison with OpenAI embeddings (Section~\ref{sec:cross-model}) demonstrates that SSR generalizes across providers and that model choice substantially affects accuracy. Extending this to additional providers (Cohere \texttt{embed-v3}, open-source E5/BGE/GTE) and to higher-dimensional models (OpenAI \texttt{text-embedding-3-large}) would map the accuracy--dimensionality--cost tradeoff more completely.

\item \textbf{Mechanism of LLM variance compression}: Our cross-model control (Section~\ref{sec:crossmodel-circularity}) establishes that variance compression is a general LLM property, but does not identify the mechanism. Candidate explanations include shared pretraining corpora encoding common scale representations, deterministic decoding (temperature~$= 0$) eliminating stochastic variation, or fundamental properties of autoregressive text processing that compress ordinal judgments. Controlled experiments varying rating temperature, model size, and pretraining data overlap would help isolate the driving factor.
\end{enumerate}


\section{Conclusion}
\label{sec:conclusion}

We calibrated and validated Semantic Similarity Rating (SSR) as a methodologically independent measurement channel for mapping LLM-generated survey text to Likert-scale ratings. Through factorial experiments, we identified anchor statement quality as the dominant factor in mapping accuracy (+29~pp from naturalistic over formal language), with SSR achieving 65--77\% exact match across two embedding providers and 91--100\% within $\pm$1 on 69 test cases spanning 8 domains. A direct comparison with LLM baselines across two model families (Claude 87\%, GPT-4o 83\%) establishes a clear accuracy--independence tradeoff, with a control condition ruling out information asymmetry as the explanation.

The most consequential finding is the 4-fold variance compression in LLM-based rating ($\sigma^2 \approx 0.21$--$0.23$ vs.\ SSR's $0.87$), demonstrated through a pre-registered circularity experiment ($N = 345$) with cross-model control. The near-identical compression in within-model and cross-model conditions (ratio $= 0.93$) establishes this as a general property of LLM-based rating: \textit{any} LLM measurement pipeline will produce artificially precise ratings that underestimate measurement uncertainty, regardless of model identity.

These findings indicate that measurement independence is not merely desirable but practically necessary for accurate uncertainty quantification in synthetic survey data. The 12~pp accuracy improvement from Voyage to OpenAI embeddings, without changing the mathematical framework, demonstrates that model selection is a high-leverage path for closing the accuracy gap while preserving independence.


\section*{Data Availability}

The complete calibration dataset (69 test cases across 8 domains), all 15 anchor families (75 anchor statements), experiment scripts, raw results from all experiments (including the 345-text circularity experiment, cross-model circularity control, information asymmetry control condition, and multi-model comparison data for both embedding models and LLM baselines), and the SSR framework source code are available in the project repository (\href{https://github.com/Lalitronico/ssr-replication}{GitHub: Lalitronico/ssr-replication}). The reference ratings, generated texts, LLM ratings, SSR ratings, GPT-4o cross-model ratings, and no-question control results are provided as JSON files to enable full replication.

\section*{Acknowledgments}

The research design, experimental methodology, data interpretation, and all scientific conclusions are entirely the author's. Claude Code (Anthropic) was used as an AI programming assistant for developing the SSR framework and experiment scripts.

\bibliographystyle{apalike}
\bibliography{references}


\end{document}